\documentclass[preprint,review,12pt]{elsarticle}
\usepackage{graphicx}
\usepackage{amsmath,amssymb}
\usepackage{lineno}

\begin{document}
\begin{frontmatter}

 \author[isas,ut]{Yuto~Ichinohe\corref{cor1}}
 \ead{ichinohe@astro.isas.jaxa.jp}
 \author[isas,ut]{Yuusuke~Uchida}
 \author[isas,ut]{Shin~Watanabe}

 \author[hiroshima]{Ikumi~Edahiro}
 \author[isas]{Katsuhiro~Hayashi}
 \author[hiroshima]{Takafumi~Kawano}
 \author[hiroshima]{Masanori~Ohno}
 \author[isas]{Masayuki~Ohta}
 \author[oist]{Shin'ichiro~Takeda}

 \author[hiroshima]{Yasushi~Fukazawa}
 \author[isas,ut]{Miho~Katsuragawa}
 \author[ut]{Kazuhiro~Nakazawa}
 \author[isas]{Hirokazu~Odaka}
 \author[ste]{Hiroyasu~Tajima}
 \author[hiroshima]{Hiromitsu~Takahashi}
 \author[isas,ut]{Tadayuki~Takahashi}
 \author[riken]{Takayuki~Yuasa}

 \cortext[cor1]{Corresponding author}

 \address[isas]{Institute of Space and Astronautical Science, Japan Aerospace Exploration Agency, 3-1-1 Yoshinodai, Chuo, Sagamihara, Kanagawa 252-5210, Japan}
 \address[ut]{University of Tokyo, 7-3-1 Hongo, Bunkyo, Tokyo 113-0033, Japan}
 \address[hiroshima]{Hiroshima University, 1-3-1 Kagamiyama, Higashi-Hiroshima, Hiroshima 739-8526, Japan}
 \address[oist]{Okinawa Institute of Science and Technology Guraduate University, 1919-1 Tancha, Onna-son, Okinawa 904-0495, Japan}
 \address[ste]{Solar-Terrestrial Environment Laboratory, Nagoya University, Furo-cho, Chikusa, Nagoya, Aichi 464-8601, Japan}
 \address[riken]{Nishina Center for Accerelator-based Science, RIKEN, RIBF Building 403, 2-1 Hirosawa, Wako, Saitama 351-0198, Japan}

\title{The first demonstration of the concept of ``narrow-FOV Si/CdTe semiconductor Compton camera''}
 \begin{abstract}
 The Soft Gamma-ray Detector (SGD), to be deployed onboard the {\it ASTRO-H} satellite, has been developed to provide the highest sensitivity observations of celestial sources in the energy band of 60-600~keV by employing a detector concept which uses a Compton camera whose field-of-view is restricted by a BGO shield to a few degree (narrow-FOV Compton camera). In this concept, the background from outside the FOV can be heavily suppressed by constraining the incident direction of the gamma ray reconstructed by the Compton camera to be consistent with the narrow FOV. We, for the first time, demonstrate the validity of the concept using background data taken during the thermal vacuum test and the low-temperature environment test of the flight model of SGD on ground. We show that the measured background level is suppressed to less than 10\% by combining the event rejection using the anti-coincidence trigger of the active BGO shield and by using Compton event reconstruction techniques. More than 75\% of the signals from the field-of-view are retained against the background rejection, which clearly demonstrates the improvement of signal-to-noise ratio. The estimated effective area of 22.8~cm$^2$ meets the mission requirement even though not all of the operational parameters of the instrument have been fully optimized yet.
 \end{abstract}
\end{frontmatter}

\section{Introduction}
Compton cameras have been regarded as one of the most promising detector technologies in the MeV and sub-MeV energy band, where gamma rays interact with the detector material mainly via Compton scattering \cite{COMPTEL93,2000A&AS..143..145S,schonfelder2004,Boggs:2004gs,Boggs:2006kw,2011ApJ...738....8B,Bloser:2002ir,Kanbach:2004gr,2006PhDT.........3Z,Orito:2003ho,Orito:2004jq,2011ApJ...733...13T}. The Imaging Compton Telescope (COMPTEL; \cite{COMPTEL93,2000A&AS..143..145S}), which was flown on the {\it CGRO} satellite, was the first Compton camera in orbit and achieved the best sensitivity to date in this energy band, even though it suffered from severe in-orbit background \cite{schonfelder2004}.

In order to reduce the extremely high in-orbit background, many instruments flown in the past employed an active-shielding technique; e.g. the Hard X-ray Detector (HXD) on board {\it Suzaku} \cite{HXD,Suzaku,2007PASJ...59S...1M} or the Phoswich Detection System (PDS) on board {\it BeppoSAX} \cite{PDS, SAX}. However, although background events induced by cosmic-ray and Earth albedo charged particles are dramatically suppressed with such active shields, there still remained a significant amount of background events which are induced by e.g. Earth albedo neutrons or the gamma rays generated inside the detector itself because of in-orbit activation \cite{2009PASJ...61S..17F}.

\begin{figure}[]
 \begin{center}
  \includegraphics[width=3.0in]{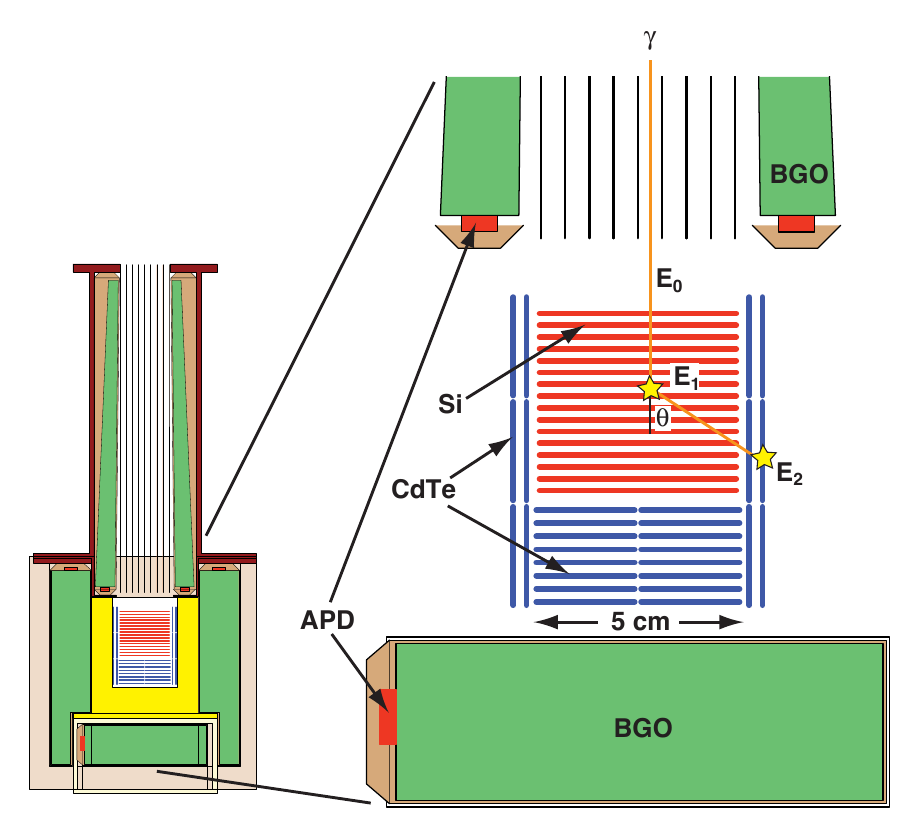}
 \end{center}
 \caption{The schematic view of SGD (color online). Compton cameras are surrounded by the active-shielding BGO scintillators. Each Compton camera consists of 32 layers of Si detectors, 8 layers of CdTe detectors and surrounding 2 layers of CdTe detectors.}
 \label{sgdschematic}
\end{figure}
The soft gamma-ray detector (SGD) \cite{Takahashi04NAR,Takahashi02-NeXT,Takahashi02,Kokubun10,Tajima05,Tajima10,Watanabe12spie,fukazawa14}, which will be on board the {\it ASTRO-H} satellite \cite{NeXT08,takahashi10,takahashi12,doi:10.1117/12.2055681}, will cover the sub-MeV energy range of 60-600~keV with the highest sensitivity ever achieved. Fig.~\ref{sgdschematic} shows a schematic view of the SGD. The SGD is based on a detector concept, which utilizes a Compton camera consisting of multiple layers of Silicon (Si) and Cadmium Telluride (CdTe) semiconductors surrounded by active-shielding, well-shaped BGO scintillators which narrowly restrict the field-of-view to about 10$^\circ\times$10$^\circ$ (narrow-FOV Si/CdTe semiconductor multilayer Compton camera; \cite{Takahashi04NAR,takahashi2001-narrowfovcc}).

The concept of the well-shaped active BGO shield has its origin in HXD, which has successfully achieved the best sensitivity in the hard X-ray band by reducing the background induced by, e.g., charged particles and gamma rays originating from out of line-of-sight directions. The main detector Compton camera complementarily reduces the background from, e.g., activation gamma rays originating from the detector itself or from neutrons originating from Earth's atmosphere, by reconstructing the direction of the detected signals.

In this paper, we, for the first time, demonstrate the concept of ``narrow-FOV Si/CdTe semiconductor multilayer Compton camera'' using the flight model (FM) of the SGD. In Section~\ref{sgdfm}, we summarize the characteristics of the SGD FM. In Section~\ref{comptonreconstruction}, we show the event reconstruction algorithm of the main detector Compton camera, and the practical methodology of the background rejection with SGD is shown in Section~\ref{methodology}. We demonstrate the concept using the FM data in Section~\ref{demonstration}.

\section{The SGD Flight Model}\label{sgdfm}
\begin{figure}[]
 \begin{center}
  \includegraphics[width=3.0in]{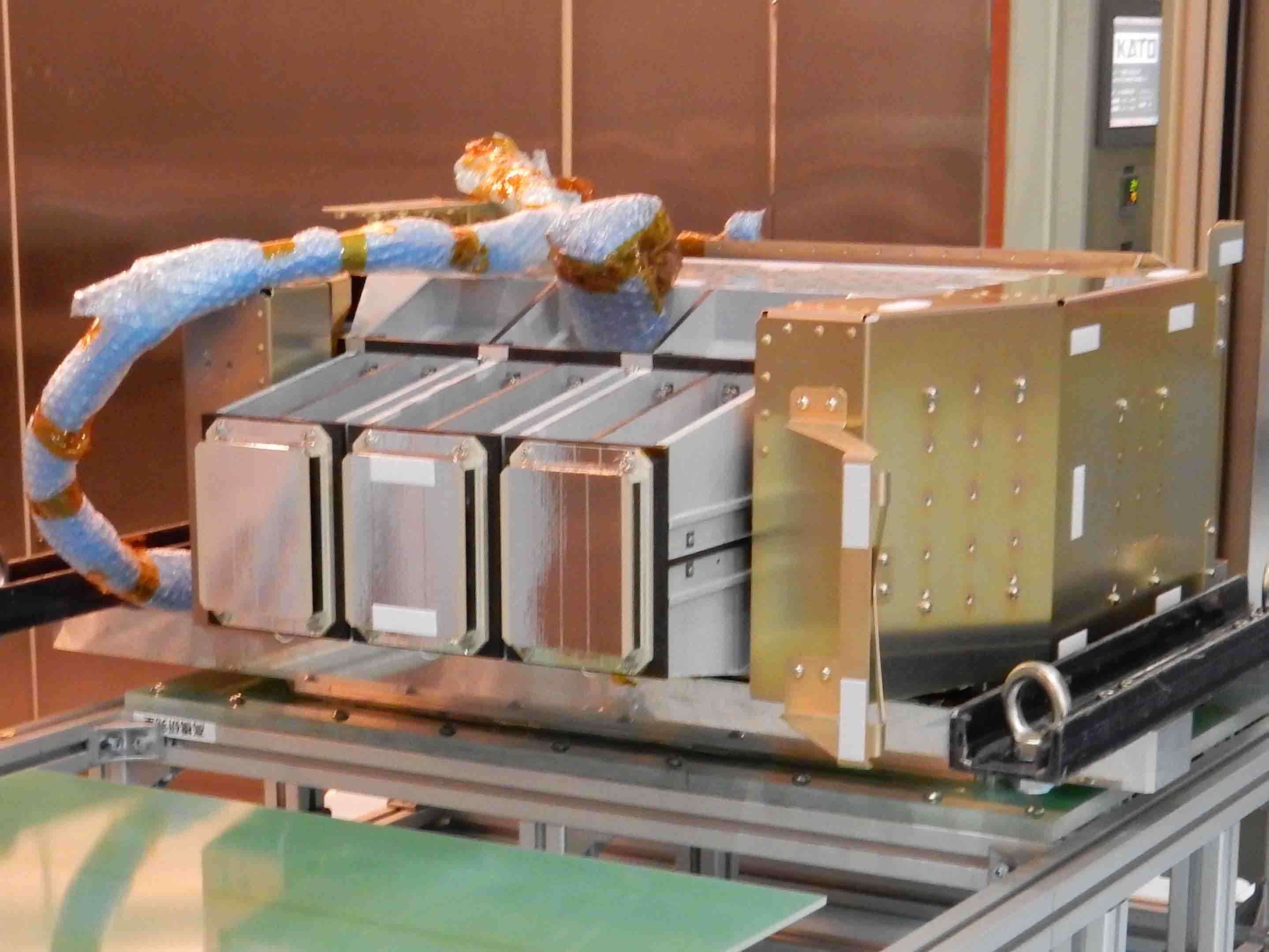}
 \end{center}
 \caption{The photo of the SGD FM (color online).}
 \label{sgdfmjpg}
\end{figure}
The SGD FM has been fabricated by integrating all the FM components after their individual tests. This FM is a result of the first complete integration of the main detector Compton cameras, the active-shielding BGO scintillators and the read-out and control electronics. The SGD FM is shown in Fig.~\ref{sgdfmjpg}.

\subsection{Main Detectors}
The detailed design and the performance overview of the main detector Si/CdTe semiconductor multilayer Compton camera are described in \cite{watanabe2014-sicdtecc}. Three identical Si/CdTe semiconductor multilayer Compton cameras constitute the main detectors of a single SGD unit. {\it ASTRO-H} will feature two such SGD units.

The Compton camera employs two types of semiconductor materials: Silicon (Si) and Cadmium Telluride (CdTe). Silicon is an ideal Compton scatterer material, since it is a low-Z element (14), and therefore its Compton scattering cross section is large and the Doppler-broadening effect is small \cite{2003SPIE.4851.1302Z}. On the other hand, Cadmium Telluride, which is high-Z (48 and 52 for Cd and Te) and high-density (5.85~g~cm$^{-3}$), has high stopping-power and is suitable as the absorber material. This combination of materials complements each other and allows for a more precise optimization between the effective area and angular resolution than that of Compton cameras made of a single material type.

Each semiconductor detector is pixelized and has a small thickness, leading to an accurate determination of the incident gamma ray's interaction within the instrument. In addition, each detector is modularized in the same size so that the loss of the effective area due to its thinness can be compensated by stacking multiple detector modules. We note that, although strip detectors could be alternatives for the pixelized detectors, we selected the pixel detectors to meet the requirement of the effective area which is crucial for astronomical observations. That is, double-sided strip detectors demand floating-readout techniques, which result in additional electronics. This leads to an increase in volume and weight of a single detector component, and thus a degradation of detection efficiency, given that the total weight and volume allowed for the SGD is determined by the entire satellite design.

One Compton camera has overall dimensions of 12~cm$\times$12~cm$\times$12~cm, consisting of 32 layers of Si pixel detector modules, 8 layers of CdTe pixel detector modules at the bottom, and surrounding 2 side layers of CdTe pixel detector modules. Each Si detector has dimensions of 51.2~mm$\times$51.2~mm$\times$0.6~mm and is pixelized to form a two-dimensional coordinate with a pixel size of 3.2~mm$\times$3.2~mm while each CdTe detector has dimensions of 25.6~mm$\times$25.6~mm$\times$0.75~mm and is also pixelized with the same pixel size of 3.2~mm$\times$3.2~mm. These Si detectors are stacked with an average pitch of 1.8~mm, whereas the CdTe detectors are stacked with an average pitches of 2.2~mm for the bottom part and 3.8~mm for the side part. These pixel sizes are optimized to have sufficient energy resolution and position determination accuracy to achieve the required angular resolution determined by the design, but to minimize the number of readout channels. The detector pitches are optimized to have sufficient number of detector layers within the given volume of the main detector. This compact configuration results in the detection efficiency (including non-diagonal components) of on-axis events about 15\% and 3\% for 100~keV and 511~keV gamma rays, respectively \cite{watanabe2014-sicdtecc}.

The signals from 13312 independent pixels included in each Compton camera are read out with 208 application specific integrated circuits (ASICs) designed for the SGD, where a set of 64 pixels are read simultaneously with a single ASIC. The Si detectors and the CdTe detectors are operated with a bias voltage of 230~V and 1000~V respectively, and the typical energy resolution is less than 2~keV at 100~keV and 1.4\% at 511 keV.

\subsection{Active Shields}
The detailed design and the signal processing methods of the BGO active shields are described in \cite{Ohno14}. Twenty-five BGO crystals, each with a thickness of about 3~cm, surround the main detectors. Thirteen of them are placed to enclose the main detectors from the sides and bottom. Four crystals are combined together to form a collimating active shield of each single Compton camera so that in total twelve of the crystals are placed at the top of the main detectors, resulting in the main detector FOV of about 10$^\circ\times$10$^\circ$.

The scintillation light of each BGO crystal is read out with an avalanche photodiode, where the size of the light-sensitive surface of each photodiode is 10~mm$\times$10~mm. The resulting energy threshold depends on the BGO shape as well as its supporting jigs, but the typical energy threshold is $\sim$150~keV, as expected from the design.

\section{Compton Camera Event Reconstruction}\label{comptonreconstruction}
In addition to rejecting background events using anti-coincidence triggers from the BGO active shield (BGO cut; in-orbit or on-ground), it is possible to further reduce the background by rejecting all Compton events whose origin is not compatible with originating from within the FOV by using the information of the gamma-ray interactions inside the Compton camera (Compton reconstruction).

Ideally, when a gamma-ray photon enters a Compton camera, it will first be scattered by a scatterer and subsequently absorbed by an absorber detector. In such a case, Compton reconstruction is a straightforward procedure because it is straightforward to calculate the first scattering angle of the gamma ray immediately using the energy deposition at the two detectors.

However, the configuration of the multilayer Compton camera employed in the SGD allows the signals induced by the multiple Compton scattering -- which are degenerated into a single signal with conventional thick detectors -- to be separated. This results in a more accurate information on the incident gamma ray's interaction inside the detector. This at the same time means that one data readout of our Compton camera may yield more than two signals.

\begin{figure}[]
 \begin{center}
  \includegraphics[width=3.0in]{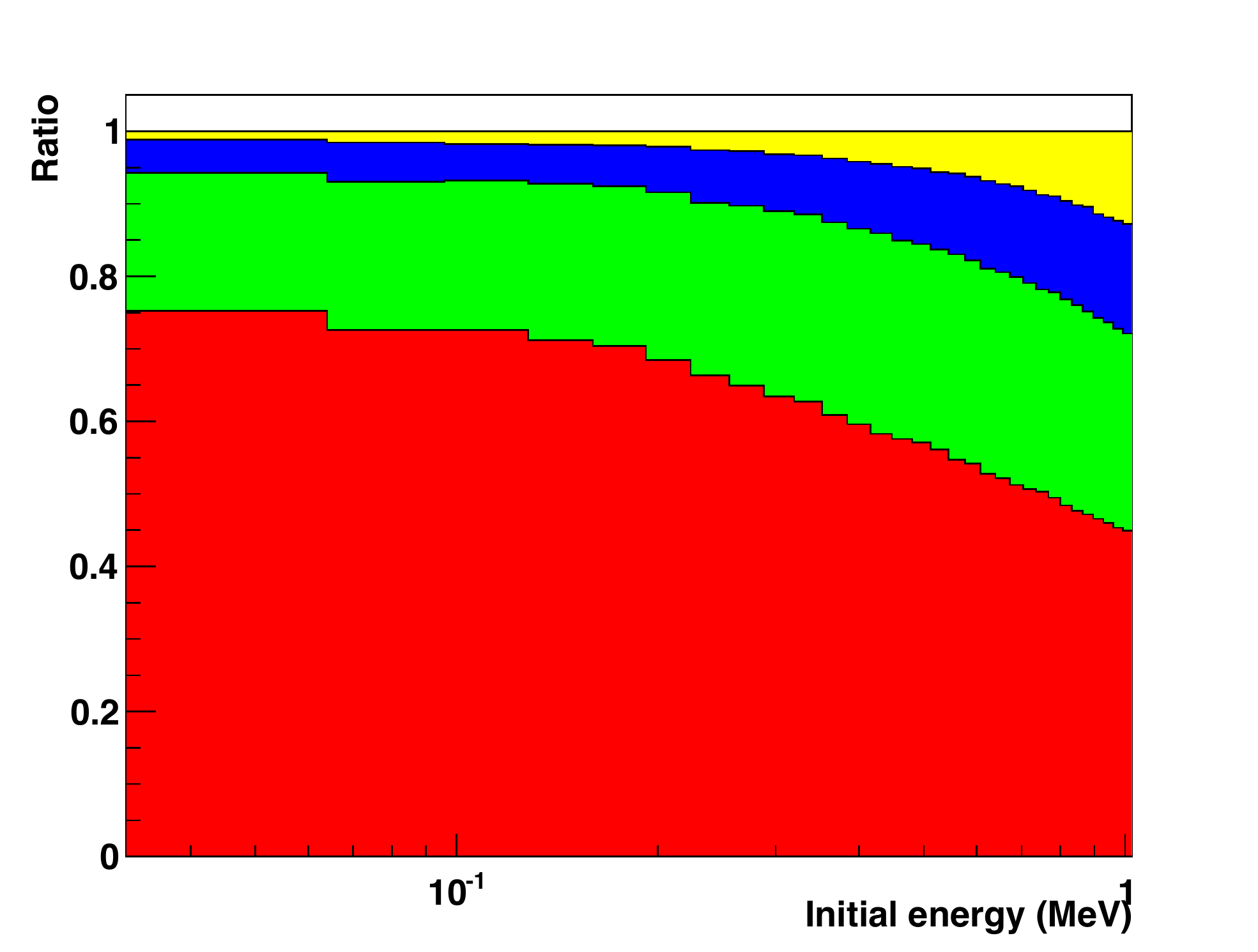}
 \end{center}
 \caption{The energy dependence of the signal multiplicity per single readout estimated with Monte Carlo simulations (color online). Red, Green, Blue and Yellow denote the signal multiplicity of two, three, four and more than four, respectively. The events interacting only once during a single readout are not plotted.}
 \label{signal_multiplicity}
\end{figure}
These signals may include not only the initial gamma-ray energy deposition, but also the energy deposition induced by other physical processes, e.g. photoabsorption of X-ray fluorescence. Fig.~\ref{signal_multiplicity} shows the energy dependence of the signal multiplicity per single readout. Even in the low-energy bands around $\lesssim$100~keV, about 30\% of the events have the signal multiplicity of more than two. Note that, the events interacting only once during a single readout are not used in the current analysis. These single-signal events represent $\sim$70\% (lowest-energy bands) to $\sim$45\% (highest-energy bands) of the total number of the events.

We developed a refined Compton reconstruction algorithm in order to maximize the event reconstruction efficiency. The algorithm comprises two steps. In the first step, the signals, which may include some spurious signals, are reduced to the hits all of which are the actual physical interaction between the incident gamma ray and the detector material\footnote{In this paper, we use the term ``hit'' to describe each physical interaction between the incident gamma ray and the detector material, whereas the term ``signal'' is used to describe the data of each readout channel no matter what the origin of the data is.}. In the second step, the order of the hits is determined.

\subsection{``Fake'' Merging}
Some of the signals may not be induced directly by the incident gamma ray, but are induced by e.g. photoabsorption of X-ray fluorescence. In such a case, if such signals are treated as an interaction (Compton scattering or photoabsorption) of the incident gamma ray, Compton reconstruction will fail. In this paper, we call these signals, which are not induced directly by the incident gamma ray, as ``fake'' signals, and these fake signals should be properly added back to its original position in order to perform Compton reconstruction correctly.

\begin{figure}[]
 \begin{center}
  \includegraphics[width=3.0in]{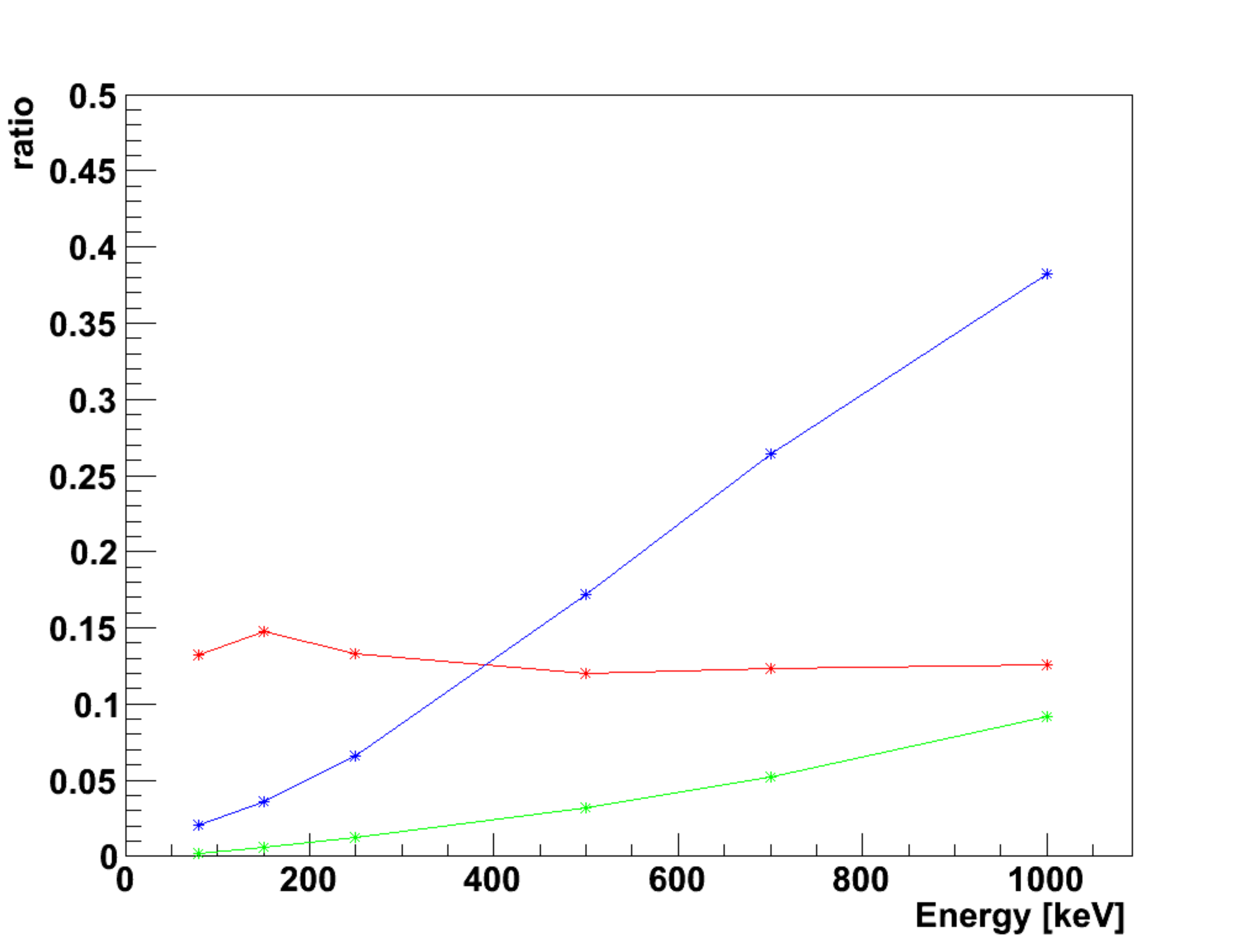}
 \end{center}
 \caption{The probability that an event includes fakes induced by photoabsorption of CdTe X-ray fluorescence (red), ionization loss of a recoiled electron (blue) or bremsstrahlung from an electron in the detector material (green), estimated with Monte Carlo simulations (color online).}
 \label{count_fake}
\end{figure}
Fig.~\ref{count_fake} shows the probability that an event includes fakes induced by photoabsorption of CdTe X-ray fluorescence (red), ionization loss of a recoiled electron\footnote{We use the term ``ionization loss of a recoiled electron'' for the signals which are caused by an electron which has escaped its original detector as well as the signals which occur in pixels which surround the pixel where the gamma-ray interaction has happened (charge sharing).} (blue) or bremsstrahlung from an electron in the detector material (green). In the case of the SGD, a fake signal is induced mostly by CdTe X-ray fluorescence or recoiled electrons.

The probability that a signal is fake highly depends on (1) the distance from the nearest signal, (2) the energy of the signal, (3) the total energy of the event, and (4) the total signal number of the event. The probability of being fake is high if the signal is located close to another signal because the fluorescence energies of Cd or Te are relatively low ($\sim$20-30~keV) and the ionization loss should be accompanied by the signals in the detector. It is also high if the energy of the signal agrees with the specific energy of CdTe X-ray fluorescence. It is again high if the total energy of the event (the initial energy of the gamma ray) is high because the probability that the recoiled electron has an energy value high enough to escape the original detector is high. Moreover, because a fake signal itself increases the total number of signals of an event, it is more possible that a fake is included in a multiple-signal event than in a two-signal event in the sense of posterior probability.

The probability of a signal being fake can be estimated with Monte Carlo simulations, but in the case of the SGD, in order to judge if a signal is fake or not, we use simpler empirical criteria which represent a smaller subspace of the whole parameter space. And if a signal is judged to be a fake, it will be added back to its nearest neighboring signal. Ultimately, Monte Carlo simulations will be performed to generate the table which associates a combination of the above mentioned parameters with the probability that a signal is fake.

\subsection{Order Determination}\label{orderdetermination}
\begin{figure}[]
 \begin{center}
  \includegraphics[width=1.8in]{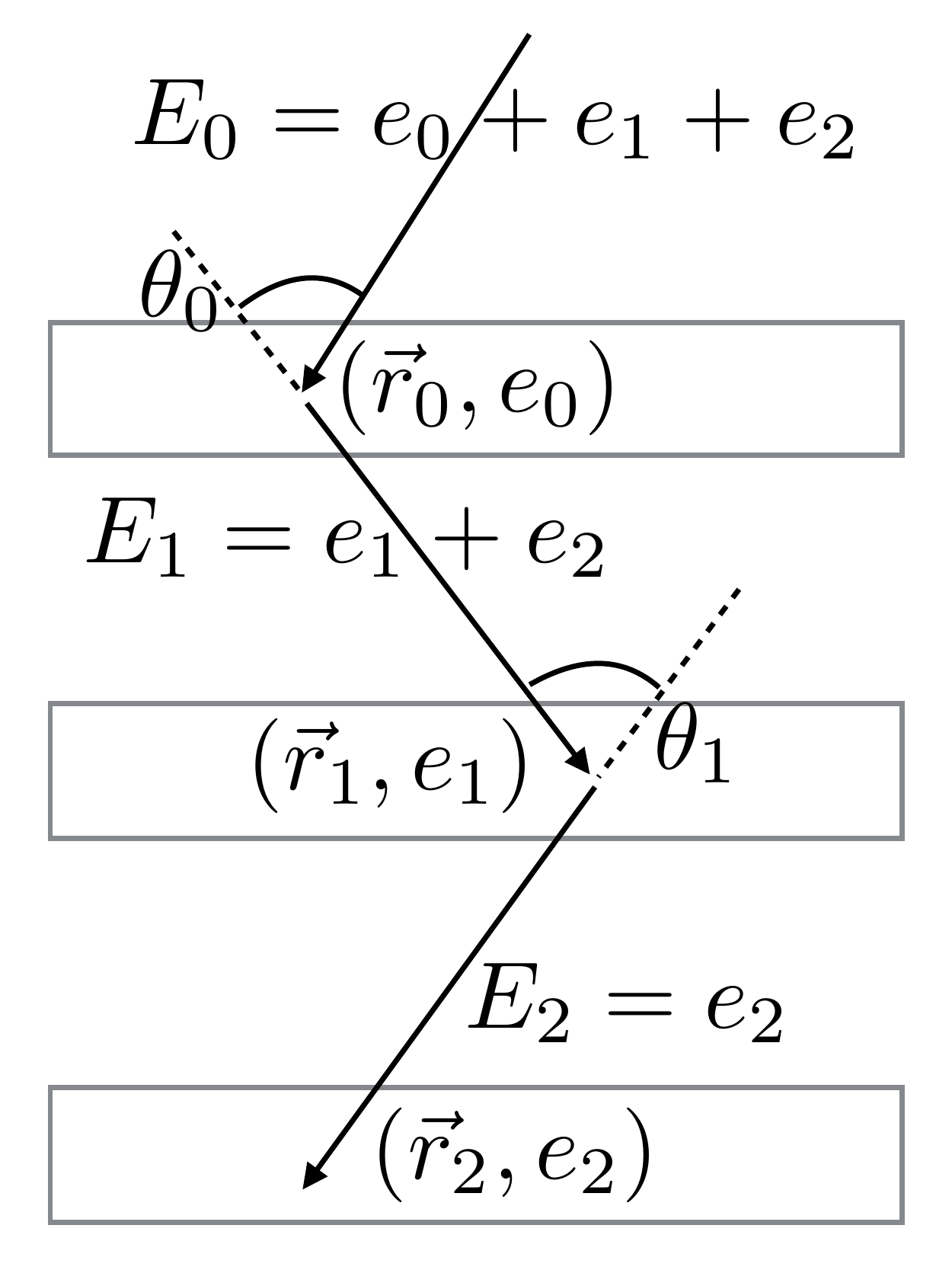}
 \end{center}
 \caption{The conceptual illustration of the order determination step in the case of $n=3$. The energy of the gamma ray before and after the first hit ($E_0=e_0+e_1+e_2$, $E_1=e_1+e_2$) are determined using the $3$ independent pieces of information of position $\vec{r}_0,\vec{r}_1,\vec{r}_2$ and energy $e_0,e_1,e_2$.}
 \label{schematic}
\end{figure}
After the signals are reduced to $n$ hits, the most plausible order of the hits is determined out of $n!$ possible sequences in the order determination step. In the beginning of this step, $n$ independent pieces of information of position $\vec{r}_i~~(0\leq i<n)$ and energy $e_i~~(0\leq i<n)$ are available. The goal is to determine which are the first and the second hits as well as to determine the energy of the gamma ray before and after the first hit ($E_0=e_0+\dots+e_{n-1}$, $E_1=e_1+\dots+e_{n-1}$). Fig.~\ref{schematic} illustrates the concept of this step.

This step is divided into three substeps. In the first substep, unphysical sequences are rejected. In the second substep, physically less probable sequences are rejected. In the third substep, the most plausible sequence is selected by tie-breaking. After each substep, if only one possible sequence is left, the sequence is selected as the most plausible order. Otherwise, the surviving sequences are passed to the next substep. In the case of the SGD, the events in which 2-4 hits are left after fake merging step are used because $>$5-hit events represent less than 1\% of the total events and are negligible.

\subsubsection{Rejecting Unphysical Sequences}
If a hit is assumed to originate from Compton scattering, the energy deposition larger than that of Compton back-scattering is prohibited. For each hit which is considered as Compton scattering, the kinematic scattering angle can be calculated using the Compton scattering formula;
\begin{eqnarray}
 \cos\theta_{iK}=1-\dfrac{m_ec^2}{E_{i+1}}+\dfrac{m_ec^2}{e_i+E_{i+1}}~~(0\leq i<n-1),\label{costhetak}
\end{eqnarray}
where $\theta_{iK}$ is the kinematic scattering angle of the $i$th hit, $e_i$ is the energy deposition of the $i$th hit, $E_{i+1}$ is the remainder of the gamma-ray energy after the $i$th hit, and $m_ec^2$ is the rest energy of an electron. Combined with the Cosine inequality, this equation is equivalent to the condition;
\begin{eqnarray}
 f_i \equiv 2E_{i+1}^2+2E_{i+1}e_i-e_im_ec^2\geq 0~~(0\leq i<n-1).\label{fi}
\end{eqnarray}

Also, the kinematic scattering angle should be consistent with the corresponding geometrical scattering angle. The geometrical scattering angle is calculated using the position of the hits;
\begin{eqnarray}
 \cos\theta_{iG}=\dfrac{\vec{r}_i-\vec{r}_{i-1}}{|\vec{r}_i-\vec{r}_{i-1}|}\cdot\dfrac{\vec{r}_{i+1}-\vec{r}_i}{|\vec{r}_{i+1}-\vec{r}_i|}~~(0<i<n-1),\label{costhetag}
\end{eqnarray}
where $\theta_{iG}$ is the geometrical scattering angle of the $i$th hit and $r_{i}$ is the position of the $i$th hit. This consistency is expressed as
\begin{eqnarray}
 g_i\equiv \cos\theta_{iG}-\cos\theta_{iK}=0~~(0<i<n-1).\label{gi}
\end{eqnarray}

If a sequence is correct, every Compton scattering candidate must satisfy both Eq.~\ref{fi} and Eq.~\ref{gi} within the errors which are propagated from the errors of the position due to the pixel sizes and the energy due to the detector energy resolution \cite{2000A&AS..145..311B,2000SPIE.4141..168O,2006PhDT.........3Z}. Conversely, if there is at least one Compton scattering candidate which doesn't satisfy the conditions above, the sequence is rejected immediately because it is unphysical. Note that, there are a few cases where an event is falsely identified as unphysical due to the energy measurement uncertainties, even when it is actually physical. However, since we incorporates the measurement uncertainty in our algorithm, the number of such events is very small and we thus ignore them.

\subsubsection{Rejecting Low Probability Sequences}
After the unphysical sequences are rejected, there may still be several possible sequences left. Although these sequences cannot be rejected immediately, the plausibility of the remaining sequences differ significantly.

For example, if an event comprises two hits, where one hit is at the Si detector and the other is at the CdTe detector, two possibility (i.e. Si-CdTe and CdTe-Si) are considered. However, since the photoabsorption cross section of Si is considerably lower than that of CdTe, the sequence where the first hit is at the Si detector and the second hit is at the CdTe detector is far more preferable than the other sequence. This is generalized to the events with more than three hits.

In the case of the SGD, we categorize the detectors into three types based on the difference in material and geometrical placement, namely, 1. Si detectors, 2. CdTe bottom detectors and 3. CdTe side detectors. Monte Carlo simulations are performed to evaluate the probability of each sequence given a combination of the types of the signals. Each sequence whose probability is lower than a tuneable threshold value (currently 10\%) is rejected as a low probability sequence. For the purpose of demonstration, we estimated the probability from the average of the probabilities separately evaluated only at a few energies, namely at 80~keV, 150~keV, 250~keV, 500~keV, 700~keV and 1000~keV, because we found that the general trend of the probability is not very sensitive to the energy.

\subsubsection{Tie-breaking}\label{tiebreaking}
After the low probability sequences are rejected, there may still be several possible sequences left. A tie-breaking is performed to single out the most plausible sequence.

In the case of the SGD, we adopt the ARM (angular resolution measure, $\Delta\theta\equiv\theta_{0K}-\theta_{0G}$) as the FOM. The ARM is defined as the difference between the first kinematic scattering angle ($\theta_{0K}$, Eq.~\ref{costhetak}) and the first geometrical scattering angle calculated assuming that the incident direction of the gamma ray is parallel to the line-of-sight direction;
\begin{eqnarray}
 \theta_{0G} = \cos^{-1}\left(-\vec{u}\cdot\dfrac{\vec{r}_1-\vec{r}_0}{|\vec{r}_1-\vec{r}_0|}\right),
\end{eqnarray}
where $\vec{u}$ is a unit vector of the direction of line-of-sight (see also Eq.~\ref{costhetag}). The sequence with the smallest ARM value is selected as the most likely sequence.

Note that, the judgement based on the calculated ARM is essentially equivalent to determining the incident direction to be the direction which is closest to the LOS direction, among many other possible directions (along the Compton cone, see also Section~\ref{armcut}). Since celestial gamma rays are expected to come from the LOS, this determination minimizes the degradation of their acceptance. On the other hand, backgrounds are expected to be more or less uniformly distributed in the whole solid angle, and thus, this determination virtually modifies the background distribution apart from the isotropic distribution to a more lopsided distribution toward the LOS direction, which may cause the bias that some of the background, which would be rejected if we chose another FOM, are not rejected.

\subsection{Treatment of Escaped Events}\label{escapedevent}
The last interaction of each incoming gamma ray is not necessarily caused by photoabsorption. A certain fraction of the incoming gamma rays end their interaction with Compton scattering then escape the main detector (escaped event).

If an event comprises more than two hits and the order of interactions are known, the energy of the gamma ray before the last interaction $E_{n-1}$ can be calculated using the energy deposition and the geometrically calculated scattering angle of the interaction second from the last $e_{n-2}$, $\theta_{(n-2)G}$ \cite{2000AIPC..510..789K,701681,2006PhDT.........3Z};
\begin{eqnarray}
 E_{n-1} = -\dfrac{e_{n-2}}{2}+\sqrt{\dfrac{e_{n-2}^2}{4}+\dfrac{e_{n-2}m_ec^2}{1-\cos\theta_{(n-2)G}}}.
\end{eqnarray}

Currently, if all sequences are rejected during the order determination step (Section~\ref{orderdetermination}) for the event with three or more hits, we treat the event to be an escaped event. We calculate $E_{n-1}$ for each possible sequence individually according to the equation above and repeat the order determination step again.

\subsection{Algorithm Demonstration}
\begin{figure}[]
 \begin{center}
  \includegraphics[width=3.0in]{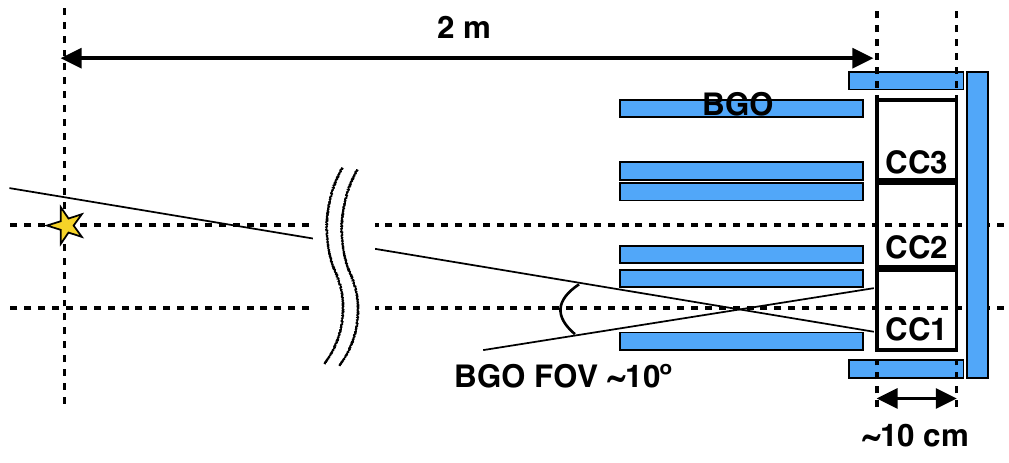}
 \end{center}
 \caption{The experimental setup of the RI irradiating experiment (color online). A $^{137}$Cs radioisotope is set in front of the Compton Camera 2 with a distance from the topmost Si pixel detector of 2~m. The FOV introduced by collimating active shields is about 10$^\circ\times$10$^\circ$.}
 \label{expschematic}
\end{figure}
We performed Compton imaging to demonstrate the reasonableness of the algorithm. Fig.~\ref{expschematic} shows the experimental setup. A $^{137}$Cs radioisotope was used for the imaging experiment. The radioisotope was set in front of Compton Camera 2 (CC2) at 2~m from the topmost Si detector. Note that, this position is not only in the BGO FOV of the CC2, but also in the BGO FOV of the CC1 (CC3).

 \begin{figure}[]
  \begin{center}
   \includegraphics[width=3.0in]{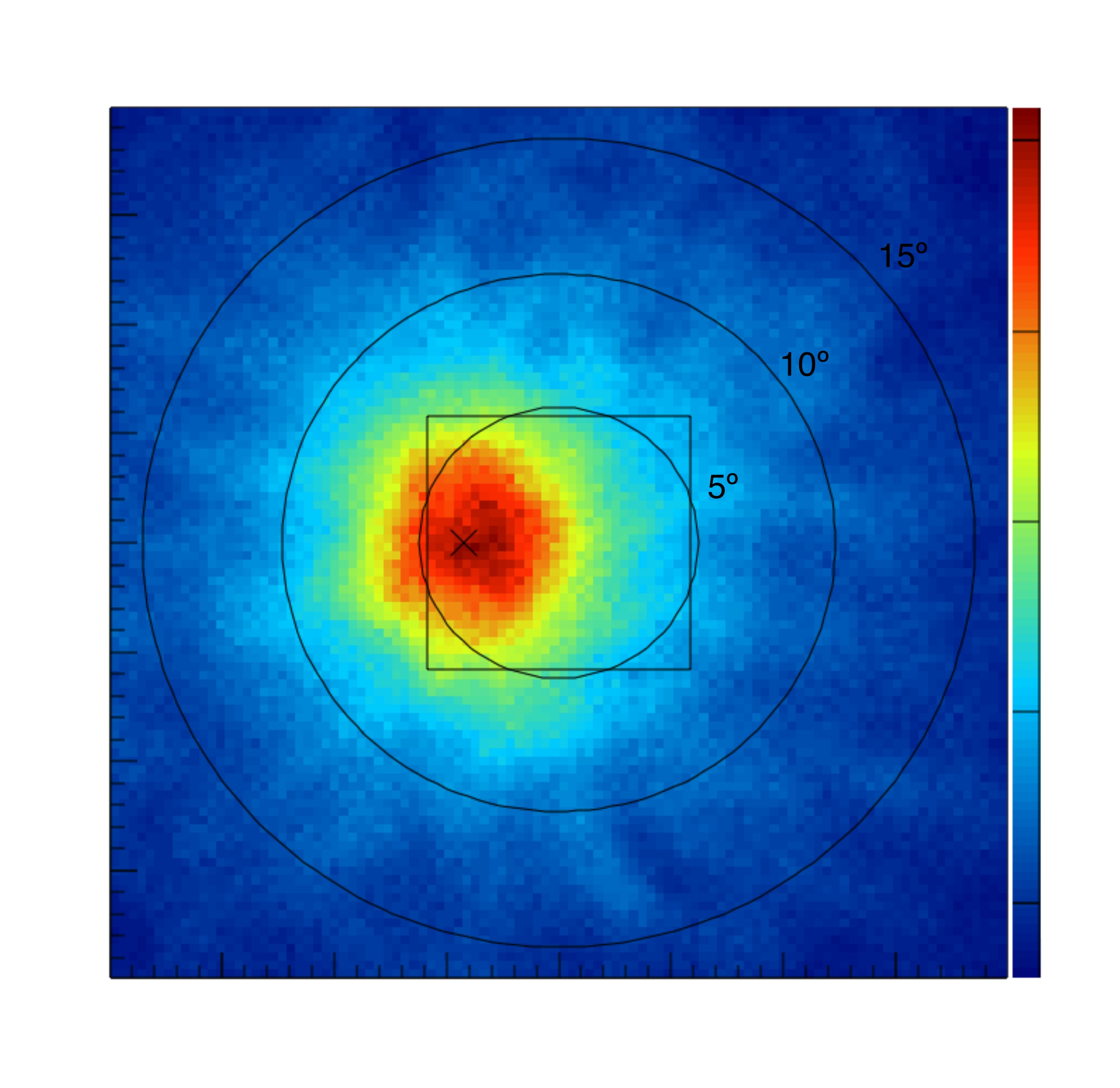}
  \end{center}
  \caption{The image of 662~keV gamma ray of $^{137}$Cs radioisotope taken with the CC1 (color online). The black square denotes the approximate FOV of the BGO active shield. The cross is the position where the radioisotope was set. The concentric circles represent the angles from the line-of-sight direction.}
  \label{csimage1}
 \end{figure}
 \begin{figure}[]
  \begin{center}
   \includegraphics[width=3.0in]{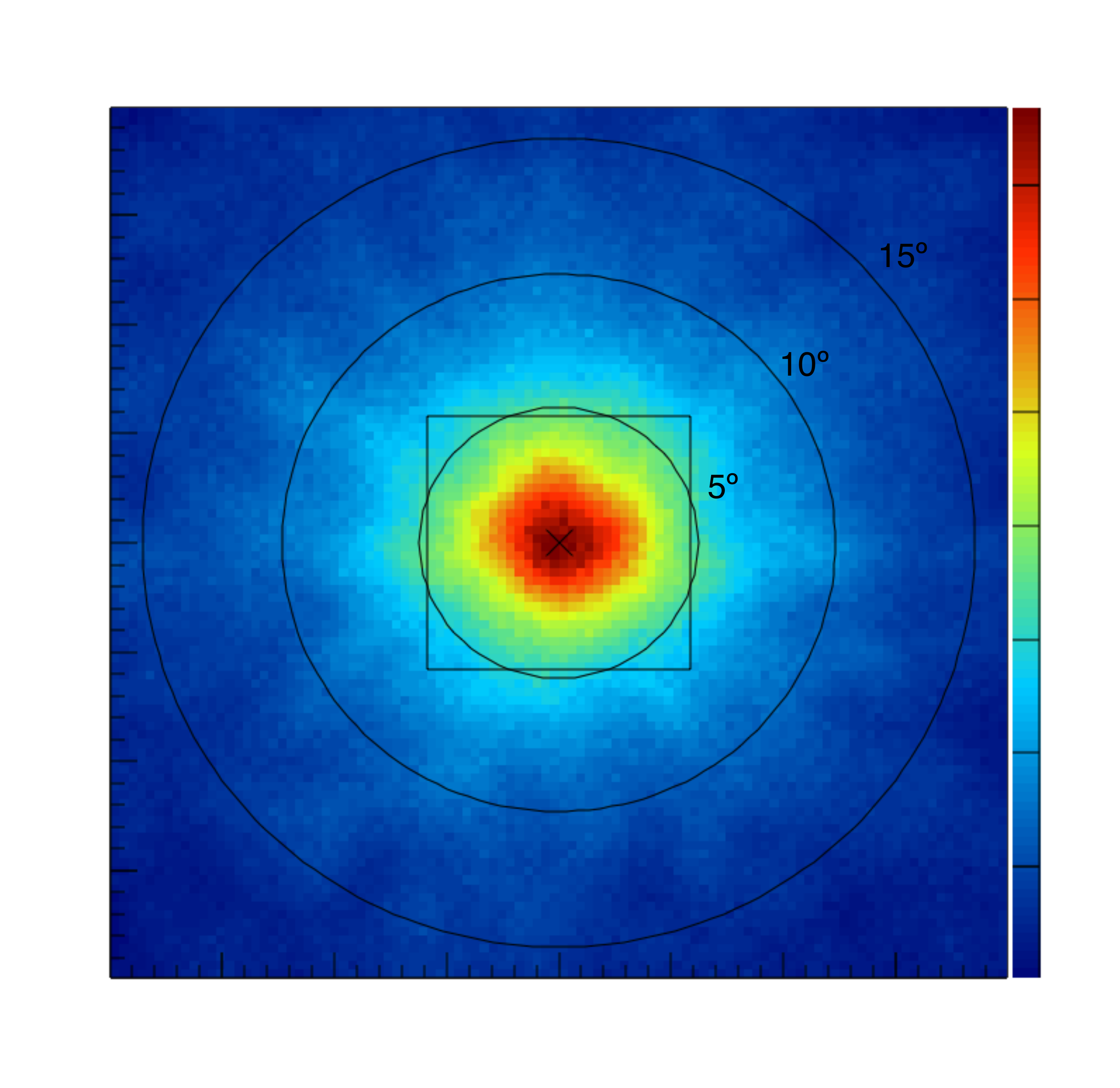}
  \end{center}
  \caption{The image of 662~keV gamma ray of $^{137}$Cs radioisotope taken with the CC2 (color online). The black square denotes the approximate FOV of the BGO active shield. The cross is the position where the radioisotope was set. The concentric circles represent the angles from the line-of-sight direction.}
  \label{csimage2}
 \end{figure}
 Fig.~\ref{csimage1} and Fig.~\ref{csimage2} show the images of 662~keV gamma ray (645-670~keV) of $^{137}$Cs radioisotope taken with the CC1 and CC2. These images are created from the reconstructed events by plotting the cross section of the Compton cones on the sky plane. The image in the sky plane is projected via equisolid angle projection (see \cite{Takeda2012859} for the detailed description on the imaging method). The black square in each image is the approximate FOV of the BGO active shield ($\sim$10$^\circ\times$10$^\circ$). Clearly the two images are different from each other. Since the radioisotope is set in front of CC2, the CC2 image forms well inside the BGO FOV. On the other hand, the CC1 image forms also inside the BGO FOV, but is displaced to the expected location. Although the image is extended because of the point spread function (PSF), the peak which represents the actual position of the target is located at the correct position with the position determination accuracy better than the PSF. This implies that the intrinsic systematic error of the algorithm is small, and thus demonstrates the reasonableness of our Compton reconstruction algorithm.

\section{Background Rejection Methodology}\label{methodology}
The SGD employs two types of background rejection techniques. One is a conventional anti-coincidence technique using BGO active shields. The other is based on the inconsistency between the incoming direction of the gamma ray and the FOV, which is specific to Compton camera technology.

\subsection{BGO cut}
Two veto signals are processed to judge whether a hit occurred in the BGO active shield during the Compton camera readout (see \cite{Ohno14} for the detailed description on BGO signal processing). One is referred to as ``Fast BGO'' and generates a quick signal to function mainly in orbit. The other is referred to as ``Hitpat'' and generates a slower but more precise signal for the purpose of ground analysis.

The Fast BGO signal aims to cancel the AD (analog-to-digital) conversions of the Compton camera in order to reduce the in-orbit deadtime due to the AD conversions, and is thus designed to be generated within 5$\mu$s from the event trigger to achieve the detector read-out cancellation. A fast digital filter is employed to generate the Fast BGO trigger. The Upper Discriminator or the Super Upper Discriminator, which employ no digital filters, also act as the Fast BGO to detect high-energy cosmic-ray events. The Fast BGO signal can also be issued without immediately canceling the AD conversions. In this case, this information will be tagged into the event data as a flag and used in the off-line analyses.

The Hitpat signal is designed to be used in off-line advanced anti-coincidence analyses. In order to achieve the lowest possible energy threshold, it employs a digital filter that is slower but more precise than that employed for the Fast BGO. The Hitpat signal is generated approximately 37$\mu$s after the event trigger, and if a Hitpat signal is issued, the information will be tagged into the event data as a flag.

\subsection{ARM cut}\label{armcut}
After determining a most plausible sequence per event following the procedure in the Section~\ref{comptonreconstruction}, the event undergoes additional, Angular Resolution Measure screening (ARM cut) and is discarded if it is unlikely to be a celestial gamma ray.

Using the energies of the gamma ray before and after the first hit ($E_0$ and $E_1$, see also Fig.~\ref{schematic}), the first scattering angle $\theta_{0K}$ of the gamma ray is calculated using the Compton scattering formula;
\begin{equation}
 \cos\theta_{0K}=1-\dfrac{m_ec^2}{E_1}+\dfrac{m_ec^2}{E_0}.
\end{equation}
Combining the calculated scattering angle with the positions of the first and the second hits ($\vec{r}_0$ and $\vec{r}_1$, see also Fig.~\ref{schematic}), the incident direction of the gamma ray $\vec{r}-\vec{r}_0$ is restricted within a conical surface (Compton cone);
\begin{equation}
 \dfrac{\vec{r}-\vec{r}_0}{|\vec{r}-\vec{r}_0|}\cdot\dfrac{\vec{r}_0-\vec{r}_1}{|\vec{r}_0-\vec{r}_1|} = \cos\theta_{0K}.\label{cone}
\end{equation}

Each Compton cone has a particular uncertainty due to the errors of the positions and the calculated energies as well as the Doppler broadening effect of the scatter material, and this error manifests itself as the width of the conical surface. If the Compton cone does not intersect the 10$^\circ\times$10$^\circ$ FOV within this error, the likelihood of the event being a celestial gamma ray is low.

The ARM value ($\Delta\theta$; see also Section~\ref{tiebreaking}), which can be interpreted as the angle between the line-of-sight direction and the cone, is a good estimator to assess whether the cone does or does not intersect the FOV. The error of the ARM $\delta[\Delta\theta]$ depends particularly on the energy of the gamma ray $E_0$ and the distance between the first two hits $|\vec{r}_0-\vec{r}_1|$, namely, the error of the ARM is smaller when the energy is higher or the distance is larger.

We performed Monte Carlo simulations, and estimated the error of the ARM as the width of the ARM distribution for each bin of the two-dimensional parameter space of $E_0$ and $|\vec{r}_0-\vec{r}_1|$. For the purpose of demonstration, we rejected the events whose ARM value is of a tuneable factor $a=1.5$ larger than its error;
\begin{equation}
 \Delta\theta > a\delta[\Delta\theta] = 1.5 \delta[\Delta\theta],
\end{equation}
as background.

We note that whether the cone does or does not intersect the FOV, depends not only on the width of the cone, but also on the shape of the FOV as well as the position of the cone vertex. Ultimately, all of these dependencies will be properly taken into consideration to estimate the likelihood.

\section{Concept Demonstration}\label{demonstration}
Here we demonstrate the concept of the narrow-FOV Si/CdTe semiconductor multilayer Compton camera using the FM of a single SGD unit. The pre-flight background data are taken during the thermal vacuum test and the low-temperature environment test. During these tests, the SGD was operated at a temperature of around $-$20$^\circ$C. The Fast BGO AD conversion canceling is set not to be invoked. All of the BGO cut background rejection is based on the off-line Fast BGO flag and the Hitpat flag.

\subsection{Background Rejection Characteristics}\label{bgrejection}
 \begin{figure*}[]
  \begin{minipage}{0.495\hsize}
   \begin{center}
    \includegraphics[width=3.0in]{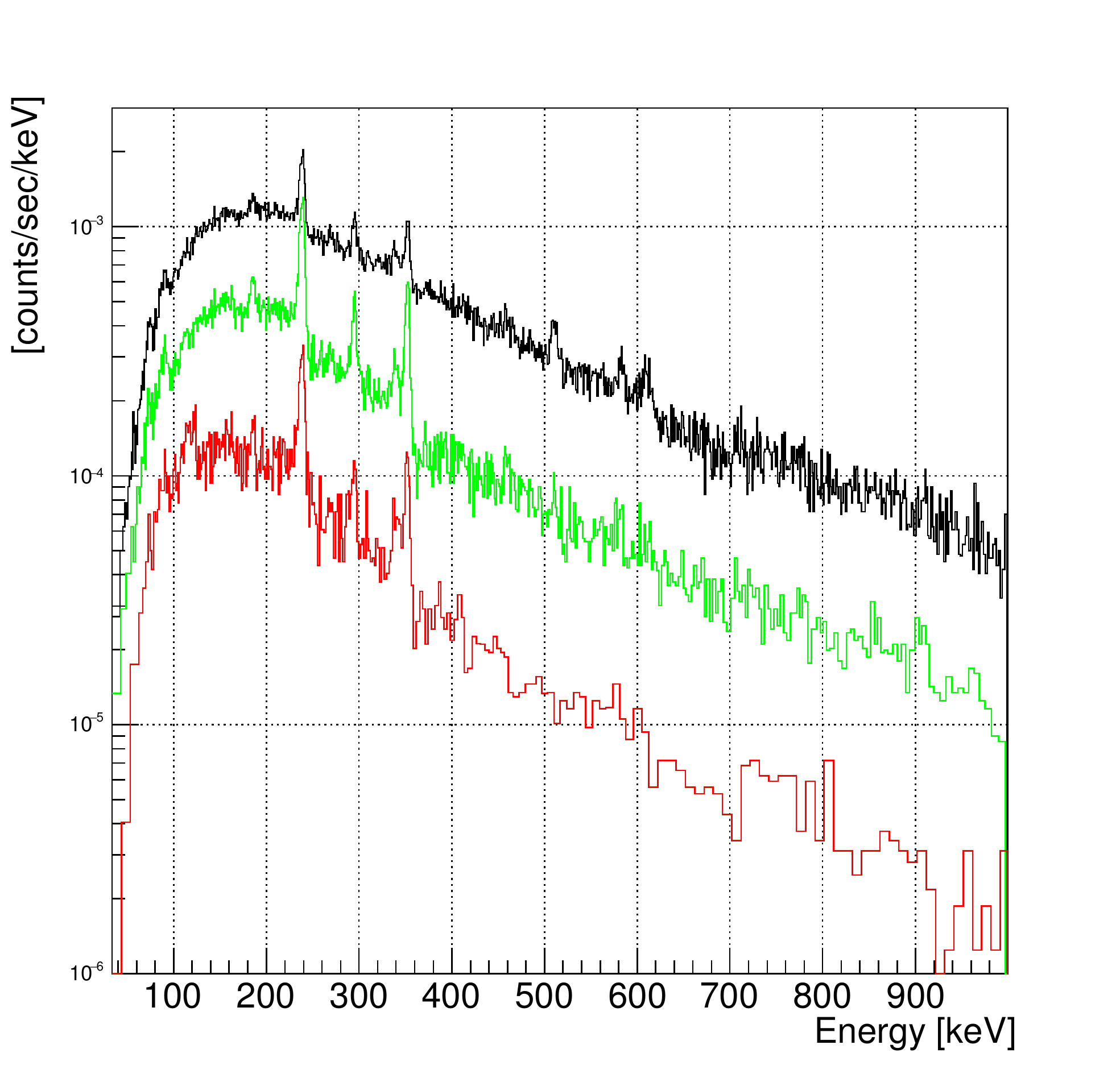}
   \end{center}
  \end{minipage}
  \begin{minipage}{0.495\hsize}
   \begin{center}
    \includegraphics[width=3.0in]{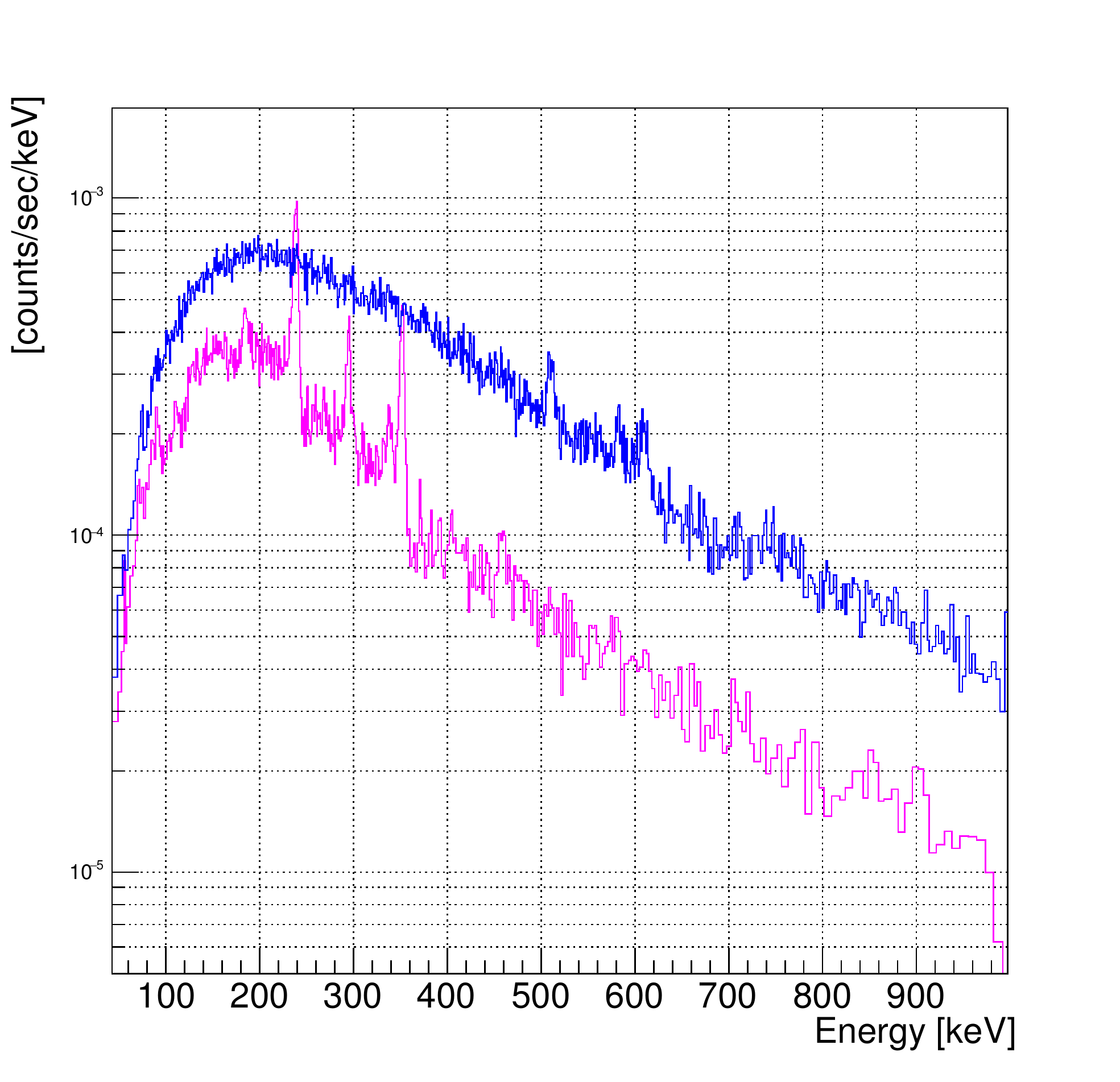}
   \end{center}
  \end{minipage}
 \caption{{\it Left:} the pre-flight background spectra taken with the CC1. The black spectrum is the raw background spectrum. The green spectrum is the background spectrum after the BGO cut background rejection is performed on the black spectrum. The red spectrum is the background spectrum after the ARM cut background rejection is performed on the green spectrum. {\it Right:} spectra made from the rejected events. The blue spectrum is the difference between the black and the green spectra in the left-hand panel. The magenta spectrum is the difference between the green and the red spectra in the left-hand panel (color online).}
 \label{bgspectra}
 \end{figure*}

 The left panel of Fig.~\ref{bgspectra} shows the actual ground background taken with the CC1. The black and the green spectra are the raw background spectrum and the background spectrum after the BGO cut rejection is performed. The difference between the black and the green spectra is shown in blue on the right-hand panel. The red spectrum is the spectrum after the ARM cut background rejection is performed on the green spectrum. The difference between the green and the red spectra is shown in magenta on the right-hand panel.

 The two spectra in the right panel of Fig.~\ref{bgspectra} show clear differences from each other. The nuclear gamma-ray lines with the energies of e.g. 511~keV (annihilation), 583~keV ($^{208}$Tl) and 609~keV ($^{214}$Bi) appear mainly on the blue (BGO cut) spectrum, while the gamma-ray lines with the energies of e.g. 239~keV ($^{212}$Pb), 295~keV ($^{214}$Pb), 339~keV ($^{228}$Ac) and 352~keV ($^{214}$Pb) appear only on the magenta (ARM cut) spectrum.

 Gamma-ray cascades occur at the decay of both $^{208}$Tl and $^{214}$Bi, resulting in multiple gamma rays emitted during a single read-out. While the energy of one of these gamma rays, which is detected in the Compton camera, is around 600~keV, the energy of the other gamma ray is 1-2~MeV. The latter gamma ray will be detected by the BGO scintillator and thus these gamma-ray lines are rejected with BGO cut.

 On the other hand, there are several lines which appear only on the ARM cut spectrum which means they cannot be rejected with the BGO cut. These lines may originate from either the outside or the inside the BGO shield. These lines do not induce gamma-ray cascades and thus cannot be rejected with active-shielding technique alone.

 These results clearly show the effectiveness of the Compton camera as main detector instead of the detectors which do not have the capability of restricting the incident gamma-ray direction. The total background counts are reduced by $\sim$70\% with the BGO cut alone, and the $\sim$70\% of the remaining background events are rejected with the ARM cut, resulting in the total background rejection efficiency of $\sim$91\%.

 \begin{figure*}[]
  \begin{minipage}{0.333\hsize}
   \begin{center}
    \includegraphics[width=2.2in]{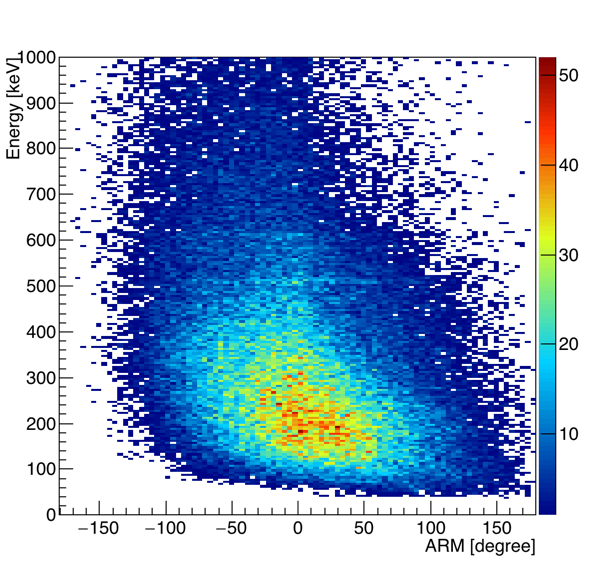}
   \end{center}
  \end{minipage}%
  \begin{minipage}{0.333\hsize}
   \begin{center}
    \includegraphics[width=2.2in]{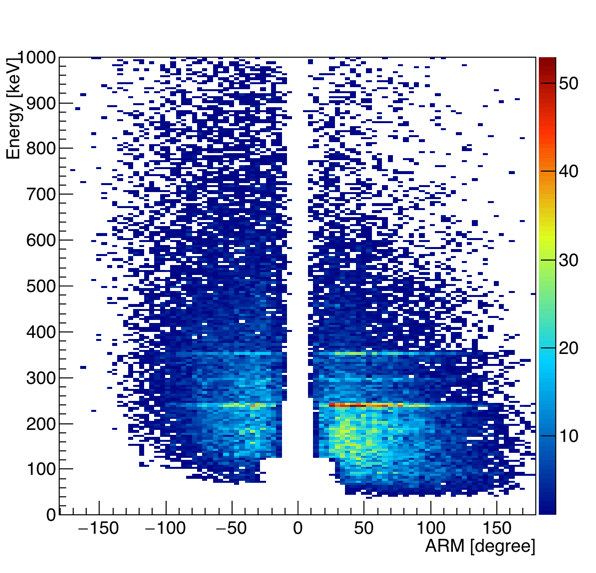}
   \end{center}
  \end{minipage}%
  \begin{minipage}{0.333\hsize}
   \begin{center}
    \includegraphics[width=2.2in]{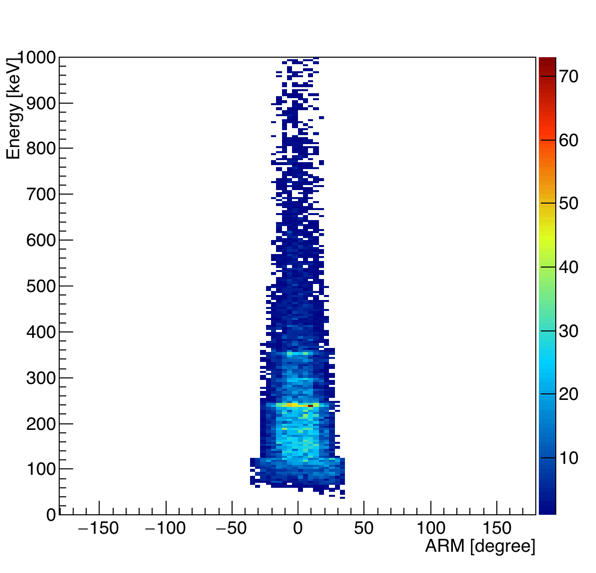}
   \end{center}
  \end{minipage}
 \caption{The correlation between the ARM $\Delta\theta$ and the calculated total energy (color online). The left-hand panel consists of the events rejected with BGO anti-coincidence (the blue spectrum in the right-hand panel of Fig.~\ref{bgspectra}). The central panel consists of the events rejected with the ARM cut (the magenta spectrum in the right-hand panel of Fig.~\ref{bgspectra}). The right-hand panel consists of the events which remain unrejected (the red spectrum in the left-hand panel of Fig.~\ref{bgspectra}).}
 \label{bgcorrelation}
 \end{figure*}
 Fig.~\ref{bgcorrelation} shows the correlations between the ARM $\Delta\theta$, which indicates the incoming direction of the gamma ray, and the calculated total energy, created using the rejected events and the events remaining unrejected. Their projection to the y-axis are the spectra shown in Fig.~\ref{bgspectra}, and the left-hand, the central and the right-hand panels correspond to the blue (BGO cut), the magenta (ARM cut) and the red spectra in Fig.~\ref{bgspectra}, respectively.

 The decreasing widths of the ARM cut with respect to the energy in the central and right-hand panels represent the fact that the error of the ARM is smaller at higher energies (see also Section \ref{armcut}). Since the ARM cut depends also on the distance between the first two hits, which is projected on the two-dimensional space of the ARM and the energy, the edges of the ARM cut are not sharp.

 The spectral lines clearly appear as a horizontal linear structures. In the central panel, the ARM of the lines which are not rejected with the BGO cut ranges over almost all angles, which means that these lines come from almost all directions. The energies of these lines are relatively low and the $\sim$3~cm thick BGO could effectively shield them if they had their origin outside the shields. Thus, the majority of the lines which are not rejected may come from inside the shield, and with the ARM cut background rejection, even these backgrounds are suppressed.

 The bright hump structure around $\Delta\theta=0$ in the left-hand panel is due to the events which are reconstructed as coming from the FOV but which generate the BGO veto signal. These events are most likely to be the escaped events (Section~\ref{escapedevent}). Generally it is difficult to recognize whether or not an event is an escaped event via using the Compton camera alone, leading to inaccurate Compton reconstruction. The BGO shields can detect such escaped events and generate the veto signal accordingly, resulting in the rejection of such inaccurately (and possibly incorrectly) reconstructed events.

 As stated above, the BGO shields and the main detector Compton camera support each other complementarily and reduce the background level to less than 10\%. The current parameters used in the Compton reconstruction and in the ARM cut are preliminary, and are to be optimized using a refined Monte Carlo simulator in which even more accurate instrumental responses such as the threshold energy or the effect of dead channels will be taken into consideration.

 \subsection{Signal Acceptance}

 Here we demonstrate that the background rejection only eliminates a small portion of the good events originating from within the FOV. We irradiated the SGD FM with a $^{137}$Cs radioisotope which is set in front of CC2 with a distance from the topmost Si pixel detector of 2~m. The experimental setup is illustrated in Fig.~\ref{expschematic}.

\begin{figure}[]
 \begin{center}
  \includegraphics[width=3.0in]{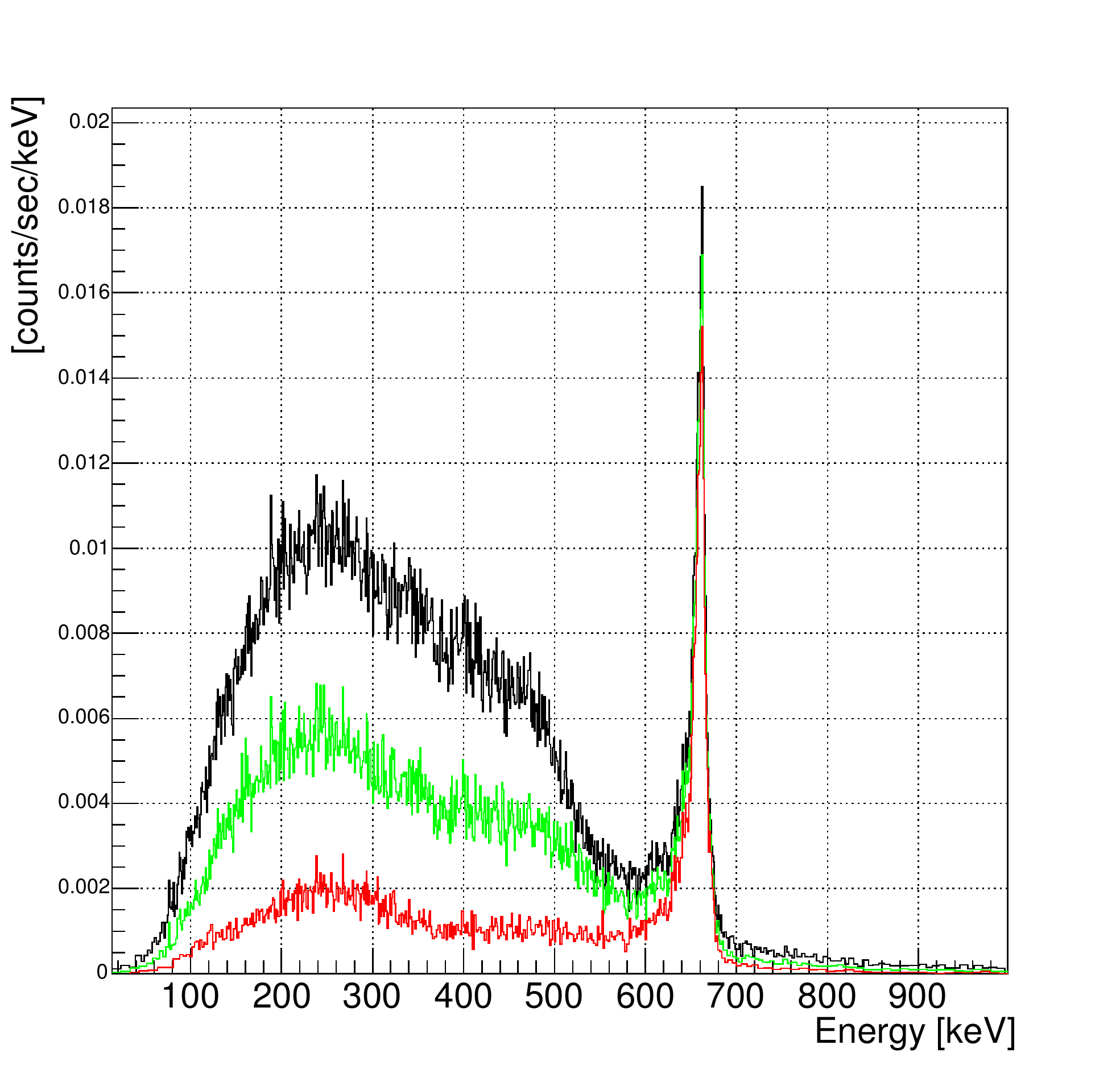}
 \end{center}
 \caption{The $^{137}$Cs spectra taken with the CC2 (color online). The black spectrum is the raw spectrum. The green spectrum is the same spectrum after the BGO cut. The red spectrum is after both the BGO cut and the ARM cut are applied.}
 \label{csspectra}
\end{figure}
Fig.~\ref{csspectra} shows the $^{137}$Cs spectra taken with the CC2. The black spectrum is the raw spectrum. The green and the red spectra are after the BGO cut and after both the BGO cut and the ARM cut are applied, respectively.

After both background rejections are applied, the spectral continuum (0-580~keV), which results in the non-diagonal component of the response, is clearly suppressed to $\sim$17\%. On the other hand, the spectral line (630-690~keV), which represents the full energy deposition of the gamma rays from the target, is suppressed only by $\sim$24\%.

\begin{figure*}[]
 \begin{minipage}{0.333\hsize}
  \begin{center}
   \includegraphics[width=2.2in]{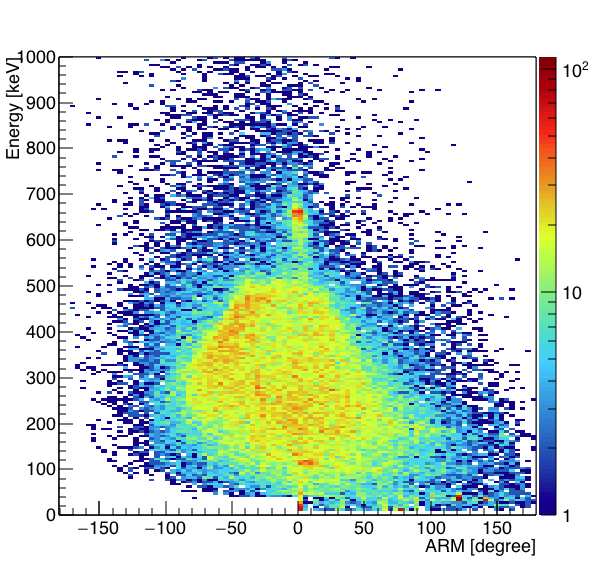}
  \end{center}
 \end{minipage}%
 \begin{minipage}{0.333\hsize}
  \begin{center}
   \includegraphics[width=2.2in]{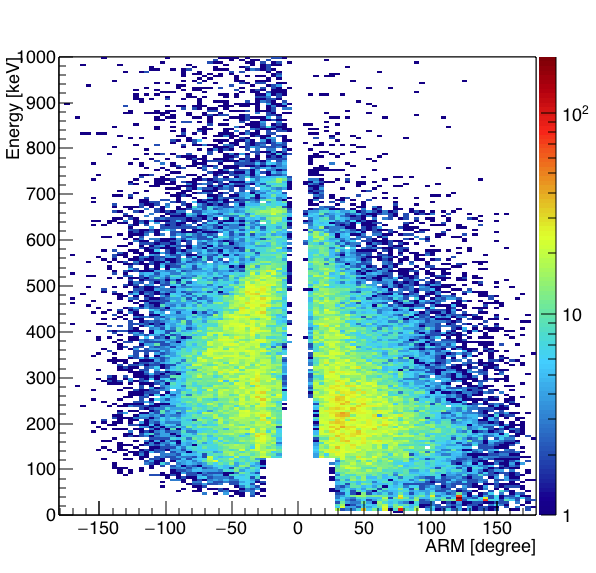}
  \end{center}
 \end{minipage}%
 \begin{minipage}{0.333\hsize}
  \begin{center}
   \includegraphics[width=2.2in]{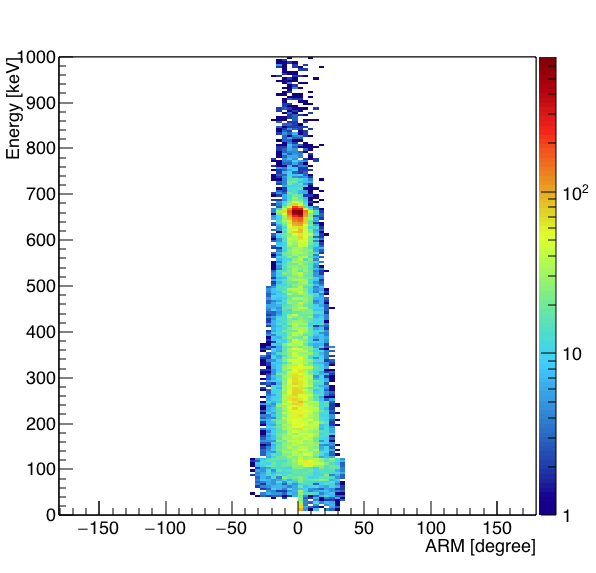}
  \end{center}
 \end{minipage}
 \caption{The correlation between the ARM $\Delta\theta$ and the calculated total energy (color online). The left-hand panel consists of the events rejected with BGO anti-coincidence (the difference between the black and the green spectra in Fig.~\ref{csspectra}). The central panel consists of the events rejected with ARM cut (the difference between the green and the red spectra in Fig.~\ref{csspectra}). The right-hand panel consists of the events which remain unrejected (the red spectrum in the left-hand panel of Fig.~\ref{csspectra}).}
 \label{cscorrelation}
\end{figure*}
Fig.~\ref{cscorrelation} shows the correlation between the ARM and the calculated total energy (see also Fig.~\ref{bgcorrelation}). Some of the 662~keV events are rejected via BGO cut. This may be caused by coincidental simultaneous events of the gamma rays from the radioisotope and the pre-flight background gamma rays. Some of the 662~keV events are also rejected with ARM cut, but it is inevitable because the Compton reconstruction algorithm is probabilistic.

$\sim$76\% of the spectral line survives the coincidental BGO cut and the probabilistic ARM cut failure. Comparing it with the background rejection efficiency (see also Section~\ref{bgrejection}) of $\sim$91\% clearly demonstrates the improvement of the signal-to-noise ratio.

\subsection{Effective Area and the Prospects for In-orbit Performance}
We estimated the effective area at 122.1~keV using a $^{57}$Co radioisotope. We irradiated the SGD FM with $^{57}$Co radioisotope which is set in front of CC1 with a distance from the topmost Si pixel detector of 2~m.

The data from the pixels which are in front of the $^{57}$Co radioisotope of all 32 layers of Si detectors are stacked and used for the effective area estimation. The total count of the pixel with the energy below 127~keV was assumed to originate from the interaction of the 122.1~keV gamma ray from $^{57}$Co radioisotope. Further assuming the approximate flatness and uniformity of the gamma rays entering the pixel, the effective area at 122.1~keV is estimated as 3.8~cm$^2$ per Compton camera.

The estimated total (6 Compton cameras) effective area at 122.1~keV is 22.8~cm$^2$. Although this is a very rough estimate, it indicates that the combination of the SGD FM and the current event reconstruction parameters meets the mission requirement on the effective area of $>$20~cm$^2$ at 100~keV.

\section{Conclusions}
We, for the first time, demonstrate the concept of a ``Narrow-FOV Si/CdTe Semiconductor Multilayer Compton Camera'' using the data taken with the SGD FM, which is the result of the first complete integration between the FOV-limiting BGO active shields and the main detector Si/CdTe multilayer Compton cameras. The pre-flight background level is suppressed to less than 10\% by the combination of the BGO cut and the ARM cut which work complementarily, while the suppression is only to $\sim$30\% with the BGO cut alone. We also demonstrate that 76\% of the signals coming from the FOV are remain not-rejected against the background rejection, which means a clear improvement of the signal-to-noise ratio.

The total effective area is estimated at 22.8~cm$^2$, which meets the mission requirement, even though the parameters on the event reconstruction are not finalized. All the parameters will be optimized, and a more accurate in-orbit performance will be investigated with the refined Monte Carlo simulator which takes the actual instrumental properties into consideration and is under development.

\section*{Acknowledgment}
We thank Dr. Greg Madejski for the English proofreading. This work was supported by JSPS KAKENHI Grant Numbers 14079207, 24105007, 24244014 and 25800151. YI is financially supported by a Grant-in-Aid for Japan Society for the Promotion of Science (JSPS) Fellows.


\section*{References}

\begin{thebibliography}{10}
\expandafter\ifx\csname url\endcsname\relax
  \def\url#1{\texttt{#1}}\fi
\expandafter\ifx\csname urlprefix\endcsname\relax\def\urlprefix{URL }\fi
\expandafter\ifx\csname href\endcsname\relax
  \def\href#1#2{#2} \def\path#1{#1}\fi

\bibitem{COMPTEL93}
{V.~Sch\"oenfelder}, H.~Aarts, K.~Bennett, H.~de~Boer, J.~Clear, W.~Collmar,
  et~al., Instrument description and performance of the imaging gamma-ray
  telescope {COMPTEL} abroad the {Compton Gamma-Ray Observatory}, Astroph. J.
  Suppl. Series 86 (1993) 657--692.

\bibitem{2000A&AS..143..145S}
V.~Sch{\"o}nfelder, K.~Bennett, J.~J. Blom, H.~Bloemen, W.~Collmar, A.~Connors,
  R.~Diehl, W.~Hermsen, A.~Iyudin, R.~M. Kippen, J.~Kn{\"o}dlseder, L.~Kuiper,
  G.~G. Lichti, M.~McConnell, D.~Morris, R.~Much, U.~Oberlack, J.~Ryan,
  G.~Stacy, H.~Steinle, A.~Strong, R.~Suleiman, R.~van Dijk, M.~Varendorff,
  C.~Winkler, O.~R. Williams, {The first COMPTEL source catalogue}, Astronomy
  and Astrophysics Supplement 143~(2) (2000) 145--179.

\bibitem{schonfelder2004}
V.~{Sch{\"o}nfelder}, {Lessons learnt from COMPTEL for future telescopes}, {New
  Astronomy Reviews} 48 (2004) 193--198.
\newblock \href {http://dx.doi.org/10.1016/j.newar.2003.11.027}
  {\path{doi:10.1016/j.newar.2003.11.027}}.

\bibitem{Boggs:2004gs}
S.~E. Boggs, W.~Coburn, D.~M. Smith, J.~D. Bowen, P.~Jean, J.~M. Kregenow,
  R.~P. Lin, P.~von Ballmoos, {Overview of the nuclear Compton telescope}, New
  Astronomy Reviews 48~(1-4) (2004) 251--255.

\bibitem{Boggs:2006kw}
S.~E. Boggs, {The Advanced Compton Telescope mission}, New Astronomy Reviews
  50~(7-8) (2006) 604--607.

\bibitem{2011ApJ...738....8B}
M.~S. Bandstra, E.~C. Bellm, S.~E. Boggs, D.~Perez-Becker, A.~Zoglauer, H.~K.
  Chang, J.~L. Chiu, J.~S. Liang, Y.~H. Chang, Z.~K. Liu, W.~C. Hung, M.~H.~A.
  Huang, S.~J. Chiang, R.~S. Run, C.~H. Lin, M.~Amman, P.~N. Luke, P.~Jean,
  P.~von Ballmoos, C.~B. Wunderer, {Detection and Imaging of the Crab Nebula
  with the Nuclear Compton Telescope}, The Astrophysical Journal 738~(1) (2011)
  8.

\bibitem{Bloser:2002ir}
P.~F. Bloser, R.~Andritschke, G.~Kanbach, V.~Sch{\"o}nfelder, F.~Schopper,
  A.~Zoglauer, {The MEGA advanced Compton telescope project}, New Astronomy
  Reviews 46~(8-10) (2002) 611--616.

\bibitem{Kanbach:2004gr}
G.~Kanbach, R.~Andritschke, F.~Schopper, V.~Sch{\"o}nfelder, A.~Zoglauer, P.~F.
  Bloser, S.~D. Hunter, J.~A. Ryan, M.~McConnell, V.~Reglero, G.~DiCocco,
  J.~Kn{\"o}dlseder, {The MEGA project}, New Astronomy Reviews 48~(1-4) (2004)
  275--280.

\bibitem{2006PhDT.........3Z}
A.~C. {Zoglauer}, {First light for the next generation of Compton and pair
  telescopes : Development of new techniques for the data analysis of combined
  Compton and pair telescopes and their application to the MEGA prototype},
  Ph.D. thesis, PhD Thesis, Garching: Max-Planck-Institut f{\"u}r
  Extraterrestrische Physik, 2006, MPE Report, No.~289 (2006).

\bibitem{Orito:2003ho}
R.~Orito, H.~Kubo, K.~Miuchi, T.~Nagayoshi, A.~Takada, T.~Tanimori, M.~Ueno, {A
  novel design of the MeV gamma-ray imaging detector with Micro-TPC}, Nuclear
  Instruments and Methods in Physics Research Section A: Accelerators,
  Spectrometers, Detectors and Associated Equipment 513~(1-2) (2003) 408--412.

\bibitem{Orito:2004jq}
R.~Orito, H.~Kubo, K.~Miuchi, T.~Nagayoshi, A.~Takada, A.~Takeda, T.~Tanimori,
  M.~Ueno, {Compton gamma-ray imaging detector with electron tracking}, Nuclear
  Instruments and Methods in Physics Research Section A: Accelerators,
  Spectrometers, Detectors and Associated Equipment 525~(1-2) (2004) 107--113.

\bibitem{2011ApJ...733...13T}
A.~Takada, H.~Kubo, H.~Nishimura, K.~Ueno, K.~Hattori, S.~Kabuki, S.~Kurosawa,
  K.~Miuchi, E.~Mizuta, T.~Nagayoshi, N.~Nonaka, Y.~Okada, R.~Orito, H.~Sekiya,
  A.~Takeda, T.~Tanimori, {Observation of Diffuse Cosmic and Atmospheric Gamma
  Rays at Balloon Altitudes with an Electron-tracking Compton Camera}, The
  Astrophysical Journal 733~(1) (2011) 13.

\bibitem{HXD}
T.~Kamae, H.~Ezawa, Y.~Fukazawa, H.~M, E.~Idesawa, N.~Iyomoto, et~al.,
  {Astro-E} hard {X}-ray detector, Proc. SPIE 2806 (1996) 314.

\bibitem{Suzaku}
H.~Inoue, {Astro-E2 mission: the third X-ray observatory in the 21st century},
  Proc. SPIE 4851 (2003) 289--292.

\bibitem{2007PASJ...59S...1M}
K.~{Mitsuda}, M.~{Bautz}, H.~{Inoue}, R.~L. {Kelley}, K.~{Koyama},
  H.~{Kunieda}, K.~{Makishima}, Y.~{Ogawara}, R.~{Petre}, T.~{Takahashi},
  H.~{Tsunemi}, N.~E. {White}, N.~{Anabuki}, L.~{Angelini}, K.~{Arnaud},
  H.~{Awaki}, A.~{Bamba}, K.~{Boyce}, G.~V. {Brown}, K.-W. {Chan}, J.~{Cottam},
  T.~{Dotani}, J.~{Doty}, K.~{Ebisawa}, Y.~{Ezoe}, A.~C. {Fabian},
  E.~{Figueroa}, R.~{Fujimoto}, Y.~{Fukazawa}, T.~{Furusho}, A.~{Furuzawa},
  K.~{Gendreau}, R.~E. {Griffiths}, Y.~{Haba}, K.~{Hamaguchi}, I.~{Harrus},
  G.~{Hasinger}, I.~{Hatsukade}, K.~{Hayashida}, P.~J. {Henry}, J.~S. {Hiraga},
  S.~S. {Holt}, A.~{Hornschemeier}, J.~P. {Hughes}, U.~{Hwang}, M.~{Ishida},
  Y.~{Ishisaki}, N.~{Isobe}, M.~{Itoh}, N.~{Iyomoto}, S.~M. {Kahn}, T.~{Kamae},
  H.~{Katagiri}, J.~{Kataoka}, H.~{Katayama}, N.~{Kawai}, C.~{Kilbourne},
  K.~{Kinugasa}, S.~{Kissel}, S.~{Kitamoto}, M.~{Kohama}, T.~{Kohmura},
  M.~{Kokubun}, T.~{Kotani}, J.~{Kotoku}, A.~{Kubota}, G.~M. {Madejski},
  Y.~{Maeda}, F.~{Makino}, A.~{Markowitz}, C.~{Matsumoto}, H.~{Matsumoto},
  M.~{Matsuoka}, K.~{Matsushita}, D.~{McCammon}, T.~{Mihara}, K.~{Misaki},
  E.~{Miyata}, T.~{Mizuno}, K.~{Mori}, H.~{Mori}, M.~{Morii}, H.~{Moseley},
  K.~{Mukai}, H.~{Murakami}, T.~{Murakami}, R.~{Mushotzky}, F.~{Nagase},
  M.~{Namiki}, H.~{Negoro}, K.~{Nakazawa}, J.~A. {Nousek}, T.~{Okajima},
  Y.~{Ogasaka}, T.~{Ohashi}, T.~{Oshima}, N.~{Ota}, M.~{Ozaki}, H.~{Ozawa},
  A.~N. {Parmar}, W.~D. {Pence}, F.~S. {Porter}, J.~N. {Reeves}, G.~R.
  {Ricker}, I.~{Sakurai}, W.~T. {Sanders}, A.~{Senda}, P.~{Serlemitsos},
  R.~{Shibata}, Y.~{Soong}, R.~{Smith}, M.~{Suzuki}, A.~E. {Szymkowiak},
  H.~{Takahashi}, T.~{Tamagawa}, K.~{Tamura}, T.~{Tamura}, Y.~{Tanaka},
  M.~{Tashiro}, Y.~{Tawara}, Y.~{Terada}, Y.~{Terashima}, H.~{Tomida},
  K.~{Torii}, Y.~{Tsuboi}, M.~{Tsujimoto}, T.~G. {Tsuru}, M.~J.~L.~. {Turner},
  Y.~{Ueda}, S.~{Ueno}, M.~{Ueno}, S.~{Uno}, Y.~{Urata}, S.~{Watanabe},
  N.~{Yamamoto}, K.~{Yamaoka}, N.~Y. {Yamasaki}, K.~{Yamashita}, M.~{Yamauchi},
  S.~{Yamauchi}, T.~{Yaqoob}, D.~{Yonetoku}, A.~{Yoshida}, {The X-Ray
  Observatory Suzaku}, Publications of the Astronomical Society of Japan 59
  (2007) 1--7.
\newblock \href {http://dx.doi.org/10.1093/pasj/59.sp1.S1}
  {\path{doi:10.1093/pasj/59.sp1.S1}}.

\bibitem{PDS}
F.~Frontera, et~al., {The high energy X-ray experiment PDS on board the SAX
  satellite}, Advances in Space Research 11 (1991) 281--285.

\bibitem{SAX}
L.~Scarsi, {The SAX mission}, Advances in Space Research 3 (1984) 491--500.

\bibitem{2009PASJ...61S..17F}
Y.~Fukazawa, T.~Mizuno, S.~Watanabe, M.~Kokubun, H.~Takahashi, N.~Kawano,
  S.~Nishino, M.~Sasada, H.~Shirai, T.~Takahashi, Y.~Umeki, T.~Yamasaki,
  T.~Yasuda, A.~Bamba, M.~Ohno, T.~Takahashi, M.~Ushio, T.~Enoto, T.~Kitaguchi,
  K.~Makishima, K.~Nakazawa, Y.~Uehara, S.~Yamada, T.~Yuasa, N.~Isobe,
  M.~Kawaharada, T.~Tanaka, M.~S. Tashiro, Y.~Terada, K.~Yamaoka, {Modeling and
  Reproducibility of Suzaku HXD PIN/GSO Background}, Publications of the
  Astronomical Society of Japan 61~(sp1) (2009) 17--S33.

\bibitem{Takahashi04NAR}
T.~Takahashi, K.~Makishima, Y.~Fukazawa, M.~Kokubun, K.~Nakazawa, M.~Nomachi,
  H.~Tajima, M.~Tashiro, Y.~Terada, Hard x-ray and $\gamma$-ray detectors for
  the next mission, New Astronomy Reviews 48 (2004) 269--273.

\bibitem{Takahashi02-NeXT}
T.~Takahashi, T.~Kamae, K.~Makishima, Future hard {X}-ray and gamma-ray
  observations, Proc. SPIE 251 (2002) 210--213.

\bibitem{Takahashi02}
T.~Takahashi, K.~Nakazawa, T.~Kamae, H.~Tajima, Y.~Fukazawa, M.~Nomachi,
  M.~Kokubun, High resolution {CdTe} detectors for the next generation
  multi-{Compton} gamma-ray telescope, in: J.~E. Truemper, H.~D. Tananbaum
  (Eds.), X-ray and Gamma-ray Telescopes and Instruments for Astronomy, {SPIE},
  Vol. 4851, 2002, pp. 1228--1235.

\bibitem{Kokubun10}
M.~Kokubun, S.~Watanabe, K.~Nakazawa, H.~Tajima, Y.~Fukazawa, T.~Takahashi,
  J.~Kataoka, T.~Kamae, H.~Katagiri, G.~Madejski, K.~Makishima, T.~Mizuno,
  M.~Ohno, R.~Sato, H.~Takahashi, T.~Tanaka, M.~Tashiro, Y.~Terada, K.~Yamaoka,
  the HXI/SGD~team, Hard x-ray and gamma-ray detector for astro-h based on si
  and cdte imaging sensors, Nucl. Instrum. Methods A 623 (2010) 425--427.

\bibitem{Tajima05}
H.~Tajima, T.~Kamae, G.~Madejski, T.~Mitani, K.~Nakazawa, T.~Tanaka,
  T.~Takahashi, S.~Watanabe, et~al., Design and performance of the {Soft
  Gamma-ray Detector} for the {NeXT} mission, IEEE Trans. Nucl. Sci. 53 (2005)
  2749--2757.

\bibitem{Tajima10}
H.~Tajima, et~al., Soft gamma-ray detector for the {ASTRO-H} mission, Proc.
  SPIE 7732 (2010) 73216--773216--17.

\bibitem{Watanabe12spie}
S.~Watanabe, et~al., Soft gamma-ray detector for the {ASTRO-H} mission, Proc.
  SPIE 8443 (2012) 844326.

\bibitem{fukazawa14}
Y.~Fukazawa, et~al., Soft gamma-ray detector (sgd) onboard the {ASTRO-H}
  mission, Proc. SPIE 9144 (2014) 91442C.

\bibitem{NeXT08}
T.~Takahashi, et~al., The {NeXT} {X}-ray mission, new exploration {X}-ray
  telescope, Proc. SPIE 7011 (2008) 14T.

\bibitem{takahashi10}
T.~Takahashi, K.~Mitsuda, R.~L. Kelley, et~al., The {ASTRO-H} mission, Proc.
  SPIE 7732 (2010) 77320Z--77320Z--18.

\bibitem{takahashi12}
T.~Takahashi, K.~Mitsuda, R.~L. Kelley, et~al., The {ASTRO-H} x-ray
  observatory, Proc. SPIE 8443 (2012) 844312.

\bibitem{doi:10.1117/12.2055681}
T.~Takahashi, K.~Mitsuda, R.~Kelley, F.~Aharonian, H.~Akamatsu, F.~Akimoto,
  S.~Allen, N.~Anabuki, L.~Angelini, K.~Arnaud, M.~Asai, M.~Audard, H.~Awaki,
  P.~Azzarello, C.~Baluta, A.~Bamba, N.~Bando, M.~Bautz, T.~Bialas, R.~D.
  Blandford, K.~Boyce, L.~Brenneman, G.~Brown, E.~Cackett, E.~Canavan,
  M.~Chernyakova, M.~Chiao, P.~Coppi, E.~Costantini, J.~de~Plaa, J.-W. den
  Herder, M.~DiPirro, C.~Done, T.~Dotani, J.~Doty, K.~Ebisawa, T.~Enoto,
  Y.~Ezoe, A.~Fabian, C.~Ferrigno, A.~Foster, R.~Fujimoto, Y.~Fukazawa,
  S.~Funk, A.~Furuzawa, M.~Galeazzi, L.~Gallo, P.~Gandhi, K.~Gilmore,
  M.~Guainazzi, D.~Haas, Y.~Haba, K.~Hamaguchi, A.~Harayama, I.~Hatsukade,
  K.~Hayashi, T.~Hayashi, K.~Hayashida, J.~Hiraga, K.~Hirose, A.~Hornschemeier,
  A.~Hoshino, J.~Hughes, U.~Hwang, R.~Iizuka, Y.~Inoue, K.~Ishibashi,
  M.~Ishida, K.~Ishikawa, K.~Ishimura, Y.~Ishisaki, M.~Itoh, N.~Iwata,
  N.~Iyomoto, C.~Jewell, J.~Kaastra, T.~Kallman, T.~Kamae, J.~Kataoka,
  S.~Katsuda, J.~Katsuta, M.~Kawaharada, N.~Kawai, T.~Kawano, S.~Kawasaki,
  D.~Khangaluyan, C.~Kilbourne, M.~Kimball, M.~Kimura, S.~Kitamoto,
  T.~Kitayama, T.~Kohmura, M.~Kokubun, S.~Konami, T.~Kosaka, A.~Koujelev,
  K.~Koyama, H.~Krimm, A.~Kubota, H.~Kunieda, S.~LaMassa, P.~Laurent,
  F.~Lebrun, M.~Leutenegger, O.~Limousin, M.~Loewenstein, K.~Long, D.~Lumb,
  G.~Madejski, Y.~Maeda, K.~Makishima, M.~Markevitch, C.~Masters, H.~Matsumoto,
  K.~Matsushita, D.~McCammon, D.~McGuinness, B.~McNamara, J.~Miko, J.~Miller,
  E.~Miller, S.~Mineshige, K.~Minesugi, I.~Mitsuishi, T.~Miyazawa, T.~Mizuno,
  K.~Mori, H.~Mori, F.~Moroso, T.~Muench, K.~Mukai, H.~Murakami, T.~Murakami,
  R.~Mushotzky, H.~Nagano, R.~Nagino, T.~Nakagawa, H.~Nakajima, T.~Nakamori,
  S.~Nakashima, K.~Nakazawa, Y.~Namba, C.~Natsukari, Y.~Nishioka, M.~Nobukawa,
  H.~Noda, M.~Nomachi, S.~O'Dell, H.~Odaka, H.~Ogawa, M.~Ogawa, K.~Ogi,
  T.~Ohashi, M.~Ohno, M.~Ohta, T.~Okajima, T.~Okazaki, N.~Ota, M.~Ozaki,
  F.~Paerels, S.~Paltani, A.~Parmar, R.~Petre, C.~Pinto, M.~Pohl, J.~Pontius,
  F.~S. Porter, K.~Pottschmidt, B.~Ramsey, R.~Reis, C.~Reynolds, C.~Ricci,
  H.~Russell, S.~Safi-Harb, S.~Saito, S.-i. Sakai, H.~Sameshima, K.~Sato,
  R.~Sato, G.~Sato, M.~Sawada, P.~Serlemitsos, H.~Seta, Y.~Shibano, M.~Shida,
  T.~Shimada, P.~Shirron, A.~Simionescu, C.~Simmons, R.~Smith, G.~Sneiderman,
  Y.~Soong, L.~Stawarz, Y.~Sugawara, S.~Sugita, A.~Szymkowiak, H.~Tajima,
  H.~Takahashi, H.~Takahashi, S.-i. Takeda, Y.~Takei, T.~Tamagawa, K.~Tamura,
  T.~Tamura, T.~Tanaka, Y.~Tanaka, Y.~Tanaka, M.~Tashiro, Y.~Tawara, Y.~Terada,
  Y.~Terashima, F.~Tombesi, H.~Tomida, Y.~Tsuboi, M.~Tsujimoto, H.~Tsunemi,
  T.~Tsuru, H.~Uchida, H.~Uchiyama, Y.~Uchiyama, Y.~Ueda, S.~Ueda, S.~Ueno,
  S.~Uno, M.~Urry, E.~Ursino, C.~de~Vries, A.~Wada, S.~Watanabe, T.~Watanabe,
  N.~Werner, N.~White, D.~Wilkins, S.~Yamada, T.~Yamada, H.~Yamaguchi,
  K.~Yamaoka, N.~Yamasaki, M.~Yamauchi, S.~Yamauchi, T.~Yaqoob, Y.~Yatsu,
  D.~Yonetoku, A.~Yoshida, T.~Yuasa, I.~Zhuravleva, A.~Zoghbi, J.~ZuHone,
  \href{http://dx.doi.org/10.1117/12.2055681}{The astro-h x-ray astronomy
  satellite} (2014).
\newblock \href {http://dx.doi.org/10.1117/12.2055681}
  {\path{doi:10.1117/12.2055681}}.
\newline\urlprefix\url{http://dx.doi.org/10.1117/12.2055681}

\bibitem{takahashi2001-narrowfovcc}
T.~{Takahashi}, T.~{Kamae}, K.~{Makishima}, {Future Hard X-ray and Gamma-ray
  Observations}, in: H.~{Inoue}, H.~{Kunieda} (Eds.), New Century of X-ray
  Astronomy, Vol. 251 of Astronomical Society of the Pacific Conference Series,
  2001, p. 210.

\bibitem{watanabe2014-sicdtecc}
S.~Watanabe, et~al., {The Si/CdTe semiconductor Compton camera of the {ASTRO-H}
  Soft Gamma-ray Detector (SGD)}, Nuclear Inst. and Methods in Physics Research
  765 (2014) 192--201.

\bibitem{2003SPIE.4851.1302Z}
A.~{Zoglauer}, G.~{Kanbach}, {Doppler broadening as a lower limit to the
  angular resolution of next-generation Compton telescopes}, in: J.~E.
  {Truemper}, H.~D. {Tananbaum} (Eds.), X-Ray and Gamma-Ray Telescopes and
  Instruments for Astronomy., Vol. 4851 of Society of Photo-Optical
  Instrumentation Engineers (SPIE) Conference Series, 2003, pp. 1302--1309.
\newblock \href {http://dx.doi.org/10.1117/12.461177}
  {\path{doi:10.1117/12.461177}}.

\bibitem{Ohno14}
M.~Ohno, et~al., {Development and verification of signal processing system of
  BGO active shield onboard Astro-H}, Proc. SPIE 9144 (2014) 91445G.

\bibitem{2000A&AS..145..311B}
S.~E. {Boggs}, P.~{Jean}, {Event reconstruction in high resolution Compton
  telescopes}, Astronomy and Astrophysics Supplement 145 (2000) 311--321.
\newblock \href {http://arxiv.org/abs/astro-ph/0005250}
  {\path{arXiv:astro-ph/0005250}}, \href
  {http://dx.doi.org/10.1051/aas:2000107} {\path{doi:10.1051/aas:2000107}}.

\bibitem{2000SPIE.4141..168O}
U.~G. {Oberlack}, E.~{Aprile}, A.~{Curioni}, V.~{Egorov}, K.-L. {Giboni},
  {Compton scattering sequence reconstruction algorithm for the liquid xenon
  gamma-ray imaging telescope (LXeGRIT)}, in: R.~B. {James}, R.~C. {Schirato}
  (Eds.), Hard X-Ray, Gamma-Ray, and Neutron Detector Physics II, Vol. 4141 of
  Society of Photo-Optical Instrumentation Engineers (SPIE) Conference Series,
  2000, pp. 168--177.
\newblock \href {http://arxiv.org/abs/astro-ph/0012296}
  {\path{arXiv:astro-ph/0012296}}.

\bibitem{2000AIPC..510..789K}
J.~D. {Kurfess}, W.~N. {Johnson}, R.~A. {Kroeger}, B.~F. {Phlips},
  {Considerations for the next Compton telescope mission}, in: M.~L.
  {McConnell}, J.~M. {Ryan} (Eds.), American Institute of Physics Conference
  Series, Vol. 510 of American Institute of Physics Conference Series, 2000,
  pp. 789--793.
\newblock \href {http://dx.doi.org/10.1063/1.1303306}
  {\path{doi:10.1063/1.1303306}}.

\bibitem{701681}
N.~Dogan, D.~Wehe, Optimization and angular resolution calculations for a
  multiple compton scatter camera, in: Nuclear Science Symposium and Medical
  Imaging Conference, 1993., 1993 IEEE Conference Record., 1993, pp. 269--273.
\newblock \href {http://dx.doi.org/10.1109/NSSMIC.1993.701681}
  {\path{doi:10.1109/NSSMIC.1993.701681}}.

\bibitem{Takeda2012859}
S.~Takeda, Y.~Ichinohe, K.~Hagino, H.~Odaka, T.~Yuasa, S.~nosuke Ishikawa,
  T.~Fukuyama, S.~Saito, T.~Sato, G.~Sato, S.~Watanabe, M.~Kokubun,
  T.~Takahashi, M.~Yamaguchi, H.~Tajima, T.~Tanaka, K.~Nakazawa, Y.~Fukazawa,
  T.~Nakano,
  \href{http://www.sciencedirect.com/science/article/pii/S1875389212017774}{Applications
  and imaging techniques of a si/cdte compton gamma-ray camera}, Physics
  Procedia 37 (2012) 859 -- 866, proceedings of the 2nd International
  Conference on Technology and Instrumentation in Particle Physics (TIPP 2011).
\newblock \href
  {http://dx.doi.org/http://dx.doi.org/10.1016/j.phpro.2012.04.096}
  {\path{doi:http://dx.doi.org/10.1016/j.phpro.2012.04.096}}.
\newline\urlprefix\url{http://www.sciencedirect.com/science/article/pii/S1875389212017774}

\end{thebibliography}
\end{document}